\documentclass[twocolumn]{aastex63}

\renewcommand{\edit}[1]{\textcolor{red}{#1}}

\def\nustar{{\it NuSTAR\/}}
\def\beppo{{\it Beppo-SAX\/}}
\def\RXTE{{\it RXTE\/}}
\def\xmm{{\it XMM-Newton\/}}
\def\chandra{{\it Chandra\/}}
\def\flux{erg~s$^{-1}$~cm$^{-2}$}
\def\ROSAT{{\it ROSAT\/}}
\def\Suzaku{{\it Suzaku\/}}
\def\Swift{{\it Swift}}
\def\Ginga{{\it Ginga}}
\def\Hitomi{{\it Hitomi}}
\def\LOFAR{{\it LOFAR}}
\usepackage{lineno}
\graphicspath{{./}{figures/}}
\usepackage{amsmath}

\shorttitle{NuSTAR Observations of Abell 665 and Abell 2146}
\shortauthors{Rojas et al.}

\begin{document}

\title{\nustar\ Observations of Abell 665 and 2146: Constraints on Non-thermal Emission}

\author [0000-0002-8882-6426]{Randall A. Rojas Bolivar}
\affil{Department of Physics \& Astronomy,
University of Utah,
115 South 1400 East, Salt Lake City, UT 84112, USA}

\author[0000-0001-9110-2245]{Daniel R. Wik}
\affil{Department of Physics \& Astronomy,
University of Utah,
115 South 1400 East, Salt Lake City, UT 84112, USA}


\author[0000-0002-3132-8776]{Ay\c{s}eg\"{u}l T\"{u}mer}
\affil{Department of Physics \& Astronomy,
University of Utah,
115 South 1400 East, Salt Lake City, UT 84112, USA}

\author[0000-0002-9112-0184]{Fabio Gastaldello}
\affil{Istituto di Astrofisica e Fiseca Cosmica-Milano
Via Edoardo Bassini, 15
20133 Milano MI, Italy}

\author[0000-0001-7271-7340]{Julie Hlavacek-Larrondo}
\affiliation{D\'epartment de physique, Universit\'e de Montr\'eal, Montr\'eal, Quebec, H3C 3J7, Canada}

\author[0000-0003-0297-4493]{Paul Nulsen}
\affiliation{Center for Astrophysics, Harvard \& Smithsonian, Cambridge, MA 02138, USA}

\author[0000-0003-1997-0771]{Valentina Vacca}
\affil{INAF Osservatorio Astronomico di Cagliari, Via della Scienza 5, 09047 Selargius (CA), Italy}

\author[0000-0002-2114-5626]{Grzegorz Madejski}
\affil{SLAC National Accelerator Laboratory
2575 Sand Hill Rd
Menlo Park, CA 94025}

\author[0000-0001-5880-0703]{Ming Sun}
\affiliation{Department of Physics and Astronomy, University of Alabama in Huntsville, Huntsville, AL 35899, USA}

\author[0000-0003-0167-0981]{Craig L. Sarazin}
\affiliation{Department of Astronomy, University of Virginia, 530 McCormick Road, Charlottesville, VA 22904-4325, USA}

\author[0000-0003-2189-4501]{Jeremy Sanders}
\affiliation{Max-Planck-Institut f\"{u}r Extraterrestriche Physik, Giessenbachstra\ss e 1, Garching 85748, Germany}

\author[0000-0003-0939-8775]{Damiano Caprioli}
\affiliation{Department of Astronomy \& Astrophysics, The University of Chicago, Chicago, IL 60637, USA}

\author[0000-0002-1984-2932]{Brian Grefenstette}
\affiliation{Cahill Center for Astrophysics, 1216 E. California Blvd., California Institute of Technology, Pasadena, CA 91125, USA}

\author[0000-0001-5839-8590]{Niels-Jorgen Westergaard}
\affil{DTU Space, Technical University of Denmark,
Elektrovej Building 327, DK-2800
Kgs Lyngby, Denmark}

\begin{abstract}

Observations from past missions such as \RXTE\ and \beppo\ suggested the presence of inverse Compton (IC) scattering at hard X-ray energies within the intracluster medium of some massive galaxy clusters. In subsequent years, observations by, e.g., \Suzaku, and now \nustar, have not been able to confirm these detections.  We report on \nustar\ hard X-ray searches for IC emission in two massive galaxy clusters, Abell~665 and Abell~2146. To constrain the global IC flux in these two clusters, we fit global \nustar~spectra with three models: single (1T) and two-temperature (2T) models, and a 1T plus power law component (T$+$IC). The temperature components are meant to characterize the thermal ICM emission, while the power law represents the IC emission. We find that the 3--30~keV Abell 665 and 3--20~keV Abell 2146 spectra are best described by thermal emission alone, with average global temperatures of $kT = (9.15\pm 0.1)$~keV for Abell~665 and $kT = (8.29\pm 0.1)$~keV for Abell~2146. We constrain the IC flux to $F_{\rm NT} < 0.60 \times 10^{-12}$~\flux\ and $F_{\rm NT} < 0.85 \times 10^{-12}$~\flux\ (20--80 keV) for Abell 665 and Abell 2146, respectively both at the 90\% confidence level. When we couple the IC flux limits with 1.4 GHz diffuse radio data from the VLA, we set lower limits on the average magnetic field strengths of $>$0.14~$\mu$G and $>$0.011~$\mu$G for Abell~665 and Abell~2146, respectively.


\end{abstract}


\section{Introduction}
\label{sec:intro}

    Galaxy cluster mergers are the most energetic events in the Universe since the Big Bang, combining two or more clusters to masses on the order of $\sim$10$^{15}$~M$_\odot$ and releasing upwards of $10^{65}$~ergs in kinetic energy \citep{MarkevitchEnergetic}. These mergers heat the cluster plasma, and introduce shocks and turbulence into the intracluster medium (ICM). These shock fronts cause compressed magnetic field lines in the ICM which can (re)-accelerate existing relativistic particles through first order Fermi acceleration. There must be relativistic particles already present in order for this to occur because Fermi acceleration at weak merger shocks can only efficiently accelerate particles already exceeding thermal energies. That is to say, these shocks have low Mach numbers ($M_s < 3$), so their acceleration efficiency is low, thus preventing particles from being accelerated directly from the thermal pool \citep{2017JKAS...50...93K}. Such shocks lead to the production of radio relics.  
    
    Turbulence produced in the merger also (re)-accelerates electrons, leading to radio halos \citep{reacceleration}. The same accelerated relativistic electrons radiating synchrotron emission in the radio would be expected to produce non-thermal emission in the form of inverse Compton (IC) scattering of the cosmic microwave background in the hard X-ray bands. Measuring the non-thermal flux from these collisions is crucial, since not including a potential source of pressure could bias mass estimates of clusters based on hydrostatic equilibrium \citep[e.g.,][]{1993ApJ...407L..49B, 2009astro2010S.305V, 2019A&A...621A..39E}. Additionally, the ratio of the upper limit of the IC flux ($F_X$) to the radio flux ($F_R$) can provide a lower limit on the average magnetic field strength $B$ of the cluster. For a single relativistic electron, the ratio $F_R / F_X$ is the ratio of energy densities, $U$, of the fields that the electron is scattering:
\begin{equation}
\label{eq:1}
\frac{F_R}{F_X} = 
\frac{U_B}{U_{\rm CMB}} = 
\frac{B^2/8\pi}{aT^4_{\rm CMB}} \, .
\end{equation}

    Extending this to a power law energy distribution of electrons emitting IC and synchrotron emission at different energies and momenta, we obtain the following expression for the magnetic field strength required to account for both:
\begin{equation}
\label{eq:2}
B = C(p)(1+z)^{(p+5)/(p+1)} \times \\
\bigg(\frac{F_R}{F_X}\bigg)^{2/(p+1)}
\bigg(\frac{\nu_R}{\nu_X}\bigg)^{(p-1)/(p+1)} \, ,
\end{equation}

where $p$ is the index of electron distribution ($N(E) \propto E^{-p}$ and related to $\alpha$, the spectral index by $p = 2\alpha + 1)$ and $C(p)$ is a proportionality constant \citep{1979rpa..book.....R,1994hea..book.....L}.
\\
\indent IC emissions from nearby clusters, like the Coma Cluster, have been detected by \RXTE~\citep{Rephaeli665RXTE} and \beppo~\citep{FF665Beppo}. These results prompted further investigations by \Suzaku~and \Swift, but the latter attempts failed \citep{OtaReview}. A later study done by \citet{FabioComaNuStar} using \nustar~provided a less stringent upper limit restricted to the core of the cluster due to limitations of the telescope's field of view (FOV), which prevents it from capturing the entirety of the radio halo, and bright thermal component. Mosaic observations have been taken and are currently being studied to provide a more accurate limit. Several other clusters reported in \citet{RephaeliSummary} have marginal claims, however some, such as Abell 2163 \citep{2163RXTE}, have been ruled out using \nustar~\citep{rojas21}. In addition to the aforementioned Coma and Abell 2163 studies, \nustar~has provided upper limits on IC emission in the Bullet Cluster \citep{wikBullet} and Abell 523 \citep{Abell523}.

   Abell~665 (hereafter, A665, $z \sim 0.1819$~\citep{665Z}) is the only cluster in the Abell catalog to receive a richness class of 5, meaning that it contains at least 300 individual galaxies that are no fainter by 2 magnitudes of the third brightest galaxy \citep{abell}. X-ray data taken from \ROSAT~suggests that the cluster is going through a merger, composed of two similar mass clusters at core crossing \citep{gomez665merger}. The cluster is host to a giant radio halo \citep{665RadioHalo, Vacca665}. X-ray observations by \chandra\ show a possible shock upstream from the core that correlates with the radio emission and a temperature jump from 8 keV to 15 keV  \citep{665Shock, 665Shock2}. This temperature jump was later shown in \citet{665ShockDasadia} to correspond to a Mach number of $M_s \sim 3$, the second largest measured behind the Bullet Cluster. Temperature measurements done by \citet{Ginga} determined the temperature to be $kT = 8.26^{+0.95}_{-0.81}$~keV using the \Ginga~satellite. A spectral analysis performed by \citet{MillionIC} using \chandra~suggests a possible detection of IC scattering, with a non-thermal 0.6--7 keV flux of $F_{NT} \sim 4.2^{+1.4}_{-1.2} $  $\times 10^{-12}$~\flux.  \\ 
   \indent Abell~2146 (hereafter, A2146, $z\sim0.2323$~\citep{2146Z}), is a rare type of cluster in that it has two clearly observable shock fronts \citep{2146Shock} at a favorable inclination along the plane of the sky \citep{2146Radio}, in a similar manner to that of the Bullet Cluster. The merger between two roughly equal mass subclusters \citep{King2146} is estimated to have the first core passage less than 0.1 Gyr ago \citep{2146Shock}. A2146 was thought of as an anomaly among merging clusters, since prior to a study done using the VLA by \citet{2146Radio} it was thought to lack large diffuse radio emission. This study reveals a faint, extended radio structure, with the upstream shock containing a radio relic and the bowshock containing a radio halo. Measurements by \chandra~estimate the global temperature of the cluster to be $7.5\pm0.3$~keV \citep{2146Shock}. There have been no published results of non-thermal studies done on this cluster despite its ideal orientation, likely because of its faint radio halo until it was measured by \citet{2146Radio}.


    In this paper, we present three deep \nustar\ (Nuclear Spectroscopic Telescope Array; \citet{nustarcollab}) observations, one of A2146 and two of A665, in order to measure non-thermal IC fluxes and use them in conjunction with  data obtained from the VLA (Very Large Array) in the 1.4 GHz range to constraint the lower limit of the magnetic field strength. In Section~\ref{sec:observe}, we discuss our data reduction procedures. In Section~\ref{sec:analysis}, we discuss how we analyzed the data and determine which model best describes the emission from the galaxy clusters. In Section~\ref{sec:Summary}, we compare our results with previous IC searches and discuss the implications of our results. Appendix~\ref{app:BGD} provides an in depth explanation of the background characterization and background modelling for these clusters and in Appendix~\ref{app:GAIN} some information regarding an issue that we've noticed with the gain that has progressively increased as \nustar~ages.
For this paper, all errors shown are at the 90\% confidence level. 

\renewcommand{\footnotesize}{\scriptsize}

\section{Observations and Data Reduction} \label{sec:observe}

\begin{figure*}
\centering
\includegraphics[scale=0.5]{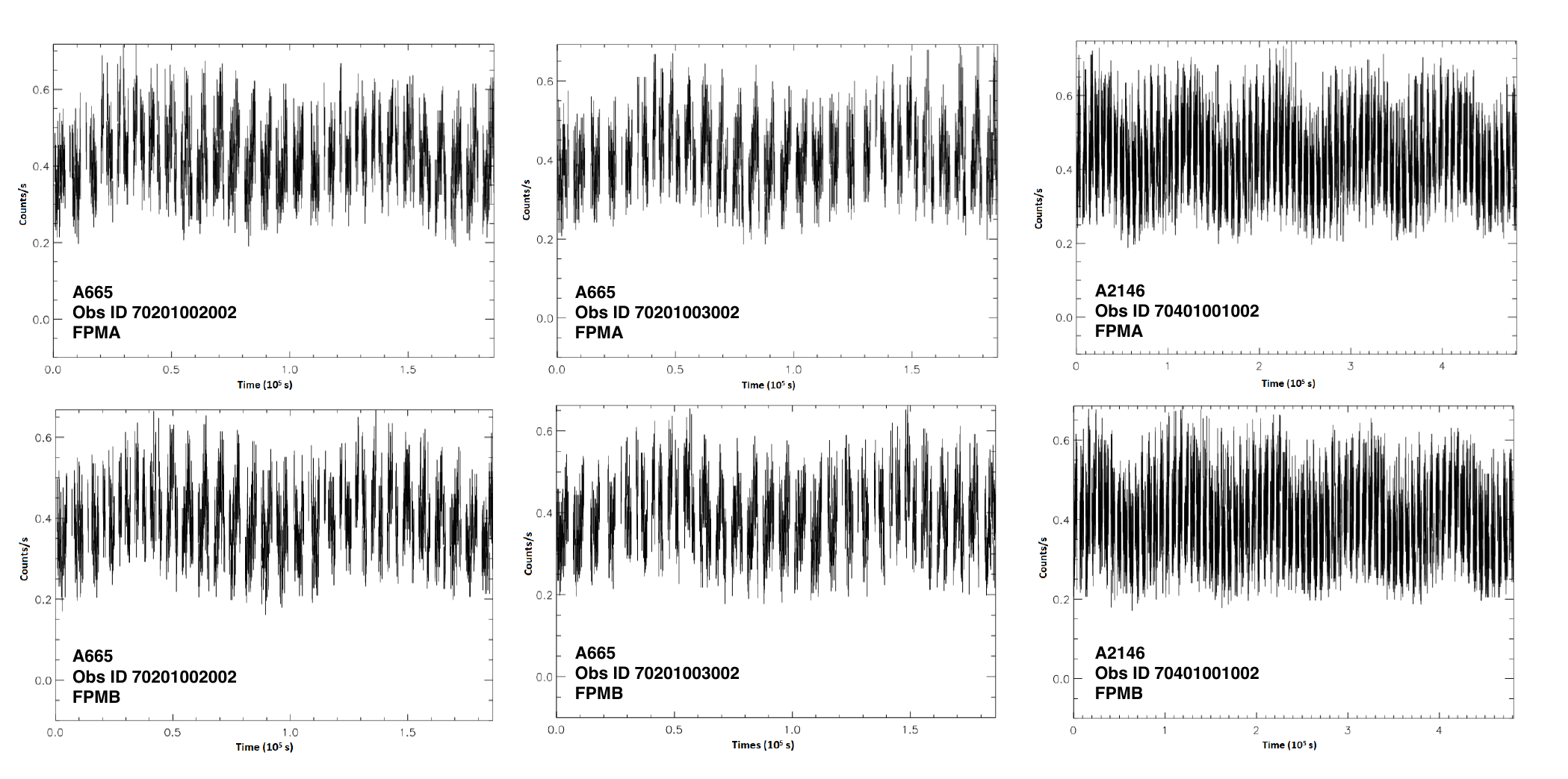}
\caption{
Filtered light curves for FPMA (top panel) and FPMB (bottom panel) telescopes following the process described in Section~\ref{sec:observe:xray}. The light curves have been filtered in the 50-160 keV energy range to eliminate SAA background contributions as well as the 1.6-20 keV range to remove solar activity background contributions. The set of light curves are in order from left to right as A665 (OBSID: 70201002002), A665 (OBSID: 70201003002), and A2146 (OBSID: 70401001002). 
\label{fig:lightcurve}
}
\end{figure*}

\subsection{X-ray} \label{sec:observe:xray}
 A665 was observed by \nustar~in segmented observations due to observation windows, the first for a total raw exposure time of 97 ks and the second, reoriented in order to better capture the shock front, for a total raw exposure time of 91 ks, including periods when the cluster was occulted by the Earth. The observations were performed between May 10th, 2017 and May 14th, 2017. The A2146 observation occurred between November 19th, 2018 and November 24th, 2018 for a total raw exposure time of 285 ks. The standard pipeline processing from {\tt HEASoft} version~6.26 and {\tt NuSTARDAS} version~1.9.4 were used to filter the data. The first step in the procedure was to remove high background periods from the data since our analysis is very sensitive to the background and any variations within it. Typically, this is achieved automatically by turning on {\tt STRICT} mode, which detects when \nustar\ has passed through the South Atlantic Anomaly (SAA), and the {\tt TENTACLE} flag, which filters for time intervals when the detectors have increased count rates when passing through the SAA, within {\tt nupipeline}.\footnote{https://heasarc.gsfc.nasa.gov/docs/nustar/analysis/nustar\_swguide.pdf}

We instead chose to manually filter our data, as this automatic filtering can be too strict and remove good data. Manual filtering is achieved by turning off the aforementioned flags and extracting good time intervals (GTIs) using light curves created for both the FPMA and FPMB telescopes by {\tt lcfilter}.\footnote{https://github.com/danielrwik/reduc} The light curves are binned in bins of 100~s and from these bins we manually find and exclude time intervals where count rates are higher than the local distribution. We repeat our exclusions three times: first to delete counts corresponding to high background due to the presence of the SAA in the 50-160 keV energy range, then again in the same energy range to more stringently reduce residual SAA contributions, and finally in the 1.6-20 keV energy range to exclude high background counts from possible solar activity. After our manual filtering, our exposure time was reduced to 91 ks in the first A665 observation, 85 ks in the second, and 255 ks in the A2146 observation. Figure~\ref{fig:lightcurve} shows our final light curves after filtering.

\begin{figure*}
\centering
\includegraphics[scale=0.6]{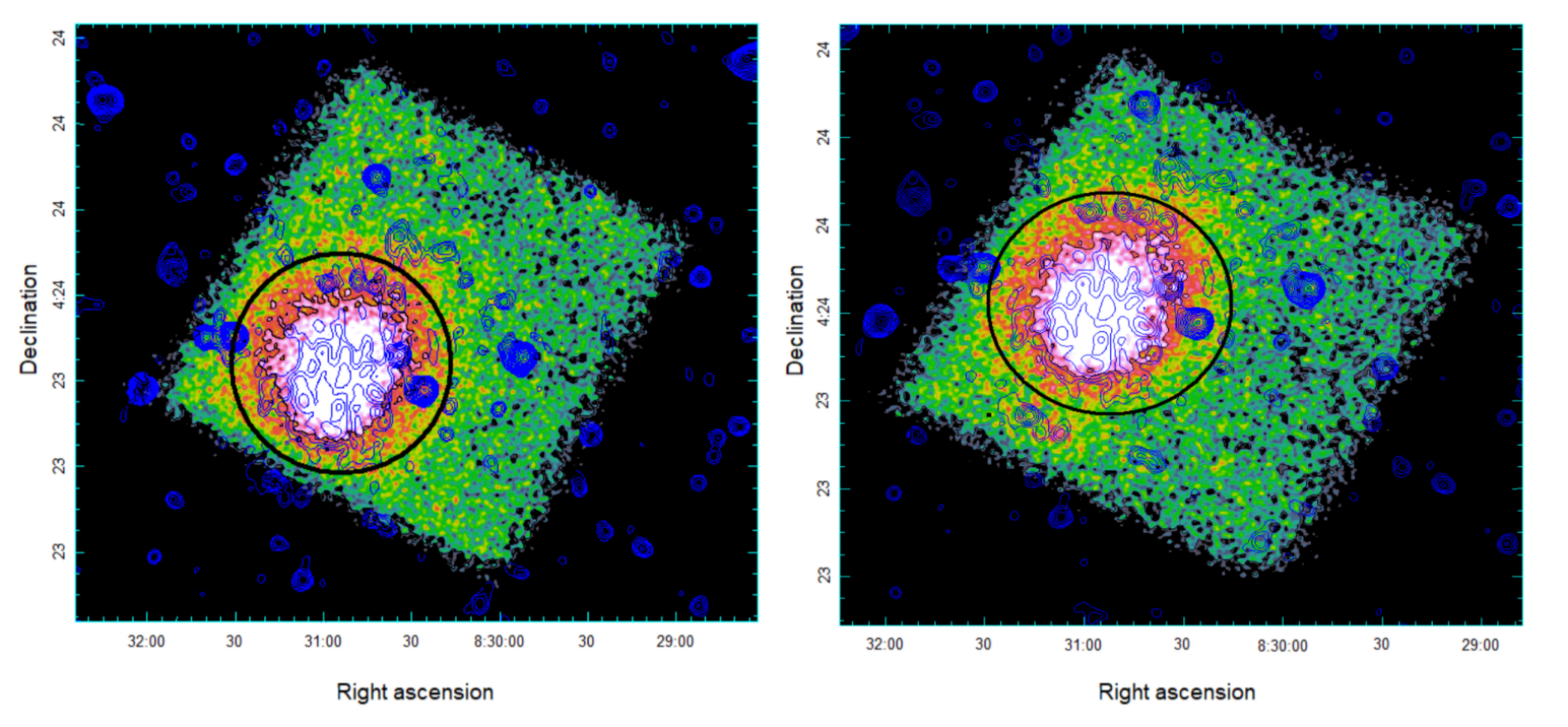}
\includegraphics[scale=0.4]{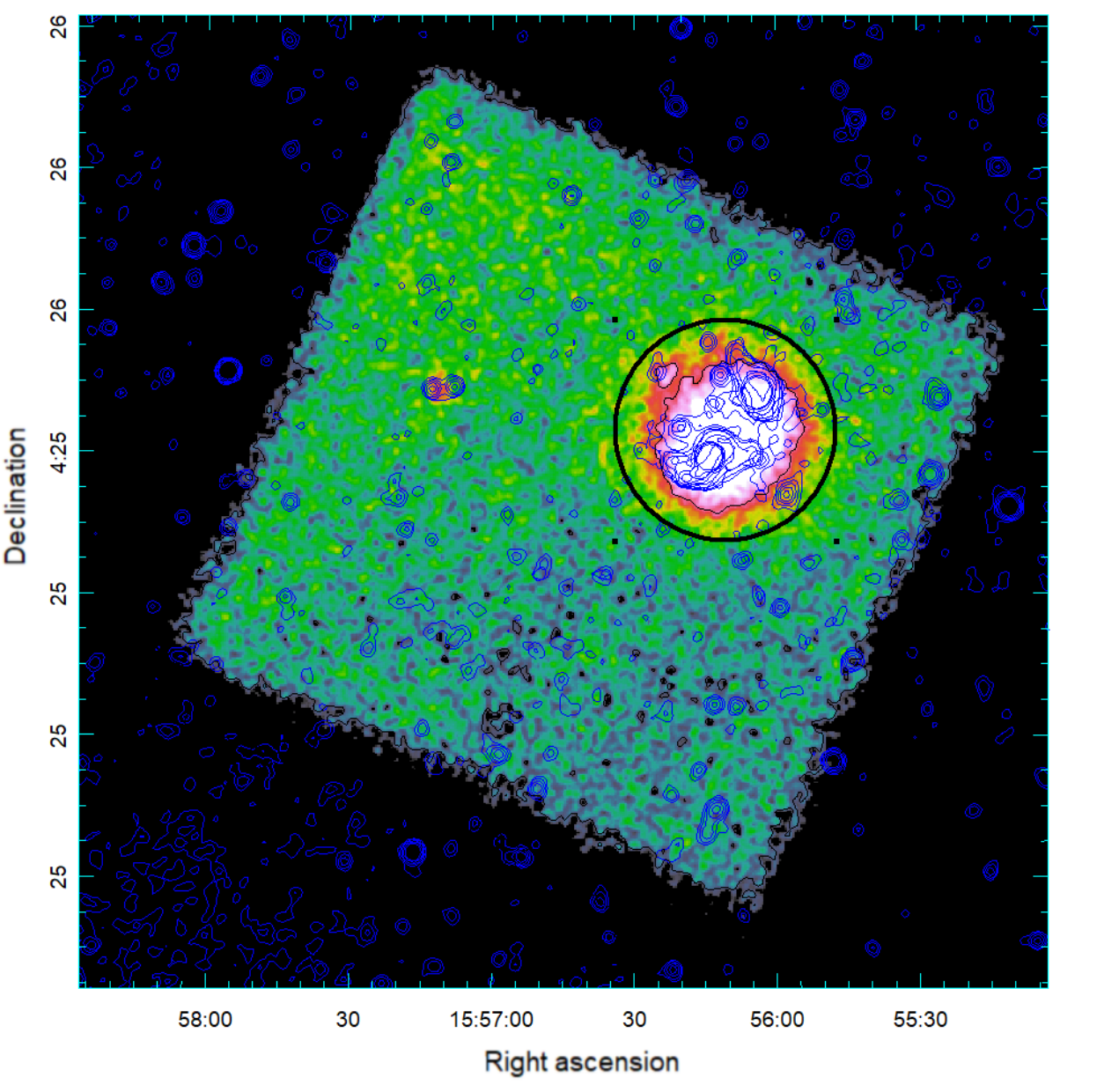}
\caption{A false color (faint to bright represented
by black to blue to green to yellow to red to white) combined
(A+B) log scaled images from 4--25~keV, smoothed by
a Gaussian kernel with $\sigma$ = 3 pix, and stretched to show
features in the outer parts of the FOV. The source regions from which spectra were extracted are shown as the black circles. Superimposed in blue are radio contours obtained from \citet{Vacca665} for A665 and \citet{2146Radio} for A2146. For A665, we display the total intensity radio contours at 1.4 GHz (combined VLA data in C and D configurations). For A2146, we display low-resolution 1-2 GHz contours. Top (left to right): A665 (OBSID: 70201002002) and A665 (OBSID: 70201003002) Bottom: A2146.} 
\label{fig:region}
\end{figure*}

The next step in our data reduction procedure was to take the newly filtered GTIs and reprocess the data using {\tt nupipeline}. Following reprocessing, we created images of the clusters using {\tt XSELECT} and produced exposure maps with {\tt nuexpomap}. Products for spectral fitting, namely the spectra, response matrices, and auxiliary response files ((PHAs, RMFs and ARFs, respectively), we produced using {\tt nuproducts}. The clean, smoothed, combined, 4-25 keV energy band images of the clusters and the source regions used for analysis in the following section are shown in Figure~\ref{fig:region}.  

 Background analysis was done following the procedures outlined in \citet{wikBullet} and \citet{rojas21}. We present the specific background modelling for these clusters in Appendix~\ref{app:BGD}.

\begin{figure*}
\centering
\includegraphics[scale=0.32]{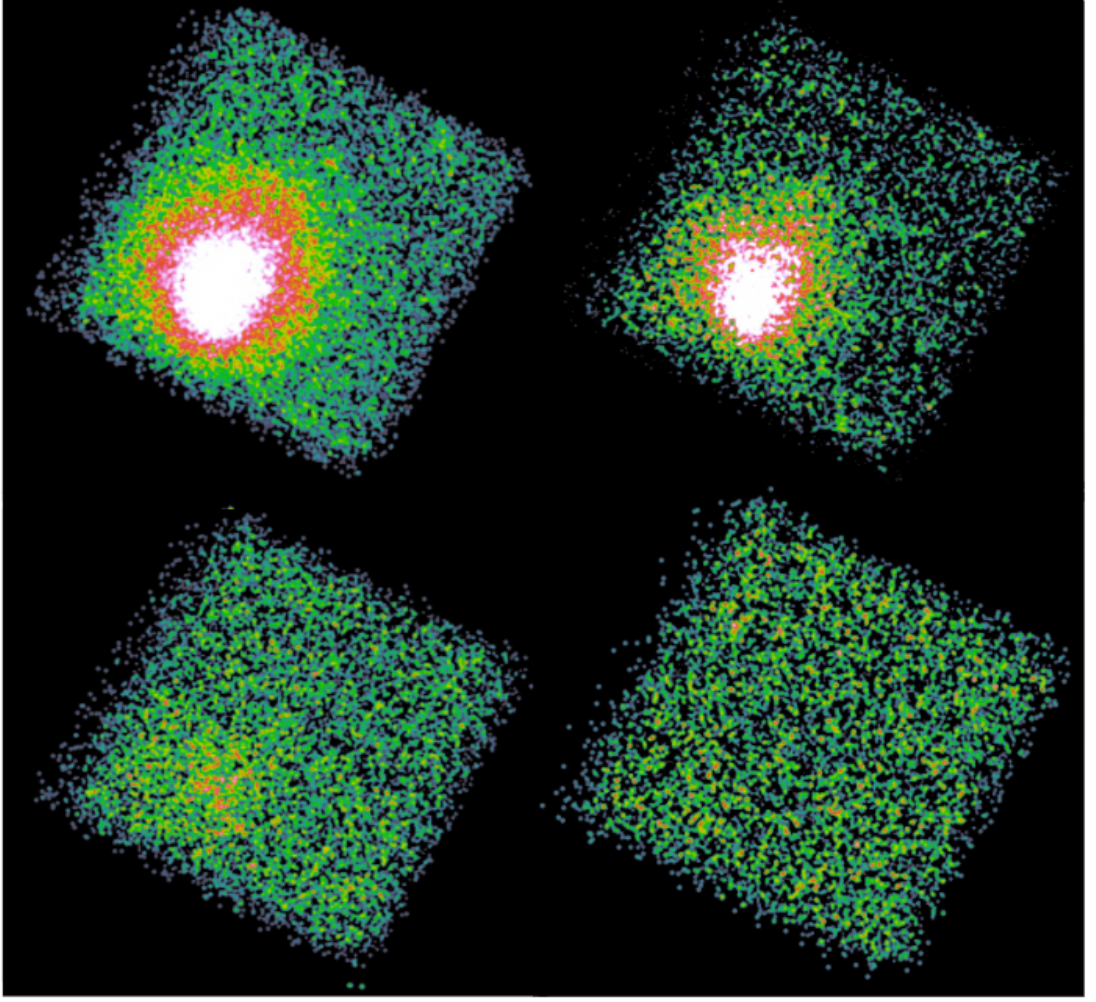}
\includegraphics[scale=0.32]{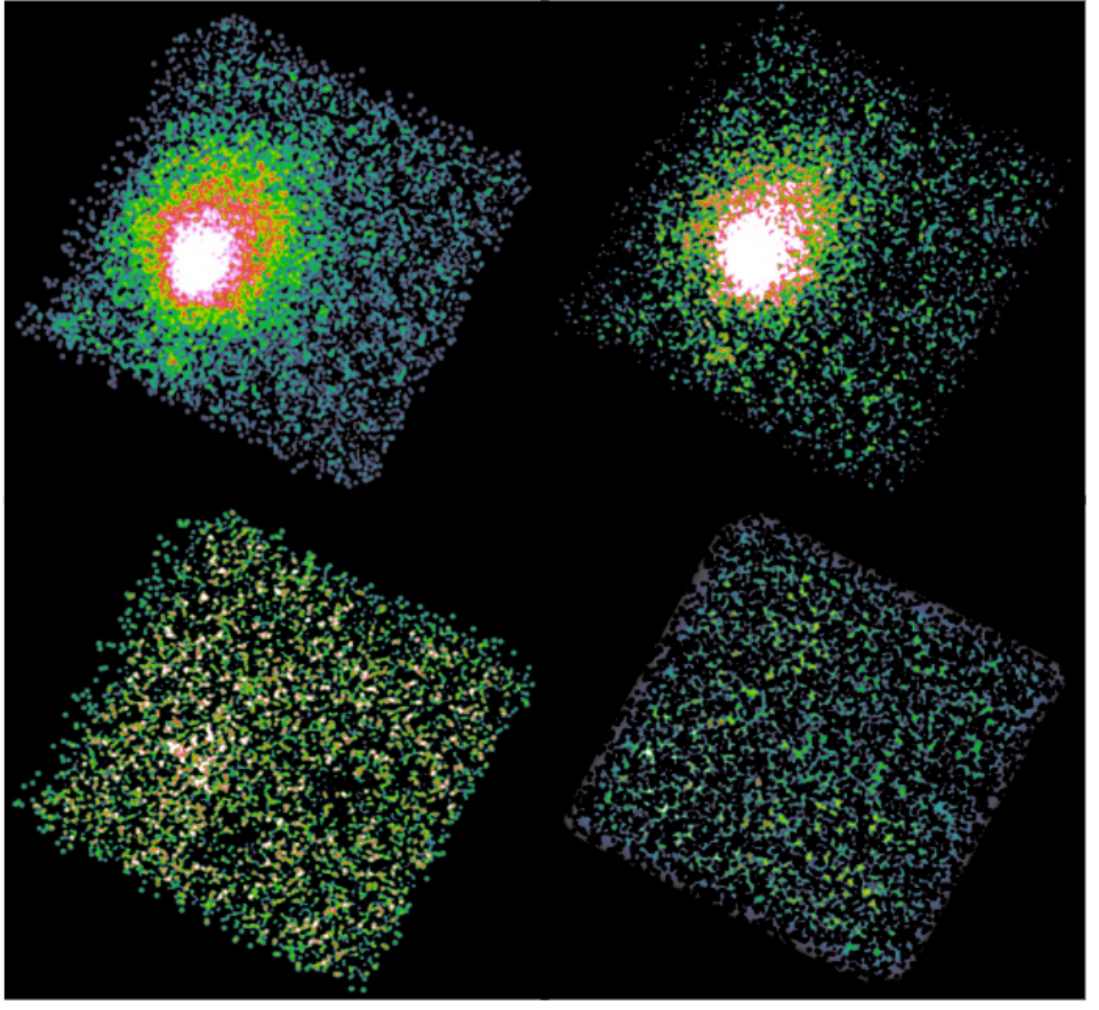}
\includegraphics[scale=0.32]{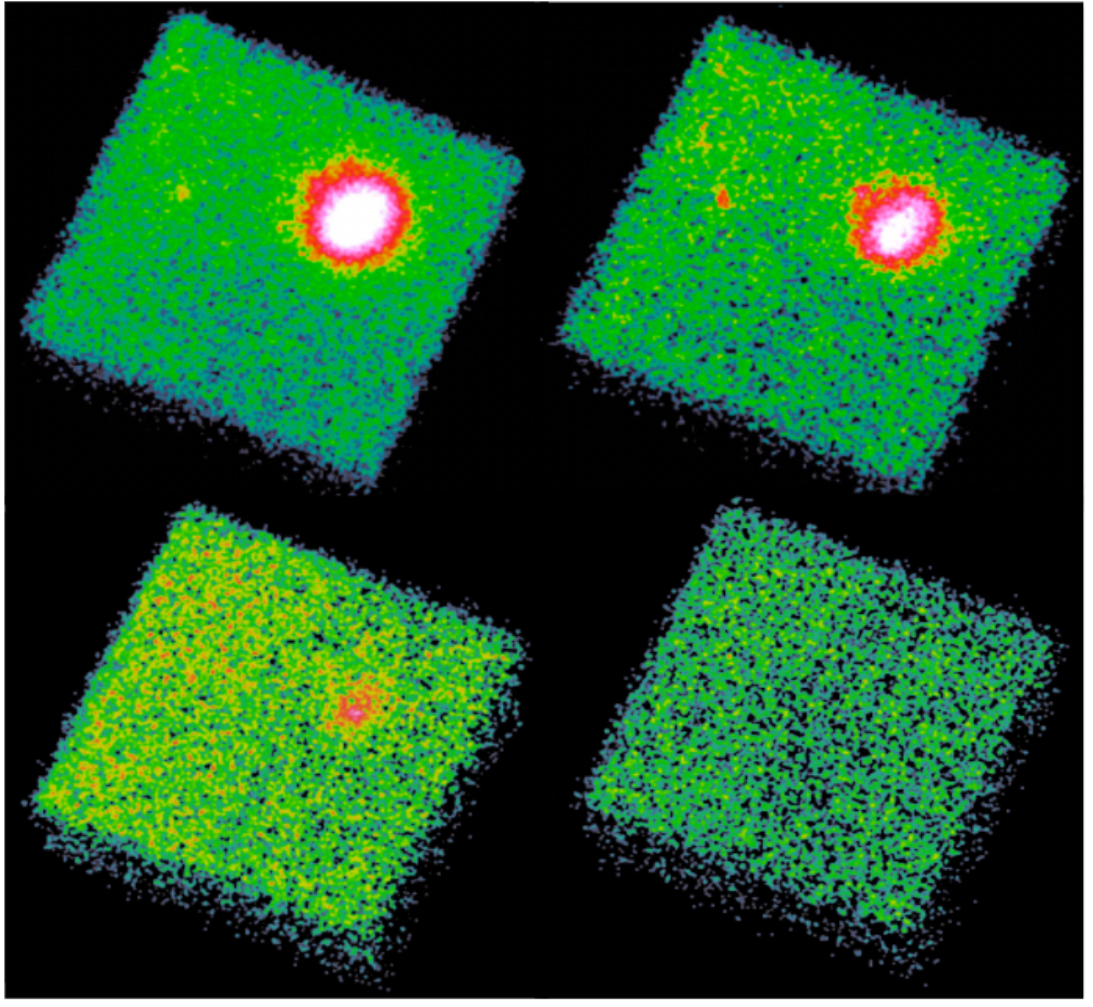}

\caption{A665 (OBSID 70201002002 on left and OBSID 70201003002 in the middle) and A2146 (right) in different energy bands. In each panel; top left: 3--8 keV; top right: 8--15 keV; bottom left: 15--30 keV; bottom right: 30--40 keV. Each image has been background subtracted and exposure corrected. They are presented in a log scale from 0 counts~s$^{-1}$~pix$^{-1}$ (in black) to 20+ counts~s$^{-1}$~pix$^{-1}$ (in white) and smoothed by a Gaussian kernel with $\sigma$ = 3~pix. 
There are fewer cluster counts present in the higher energy images and there are no obvious morphological changes with respect to the lower energy images, which are dominated by thermal photons.}
\label{fig:energybands}

\end{figure*}

\section{Analysis} \label{sec:analysis}
    In order to exclude possible point sources that might have had an effect on our spectral analysis, we generated images of the clusters in different energy bands by using {\tt xselect} to filter the PHA column. Figure~\ref{fig:energybands} shows 3-8, 8-15, 15-30, and 30-40 keV exposure-corrected, background-subtracted images. The background was modelled using the {\tt nuskybgd} routine and is discussed in detail in Appendix~\ref{app:BGD}. These images show no evidence of of major bright point sources that can contaminate our spectra.
   





\subsection{Spectra} \label{subsec:spectra}

\begin{figure*}
\centering
\includegraphics[width=183mm]{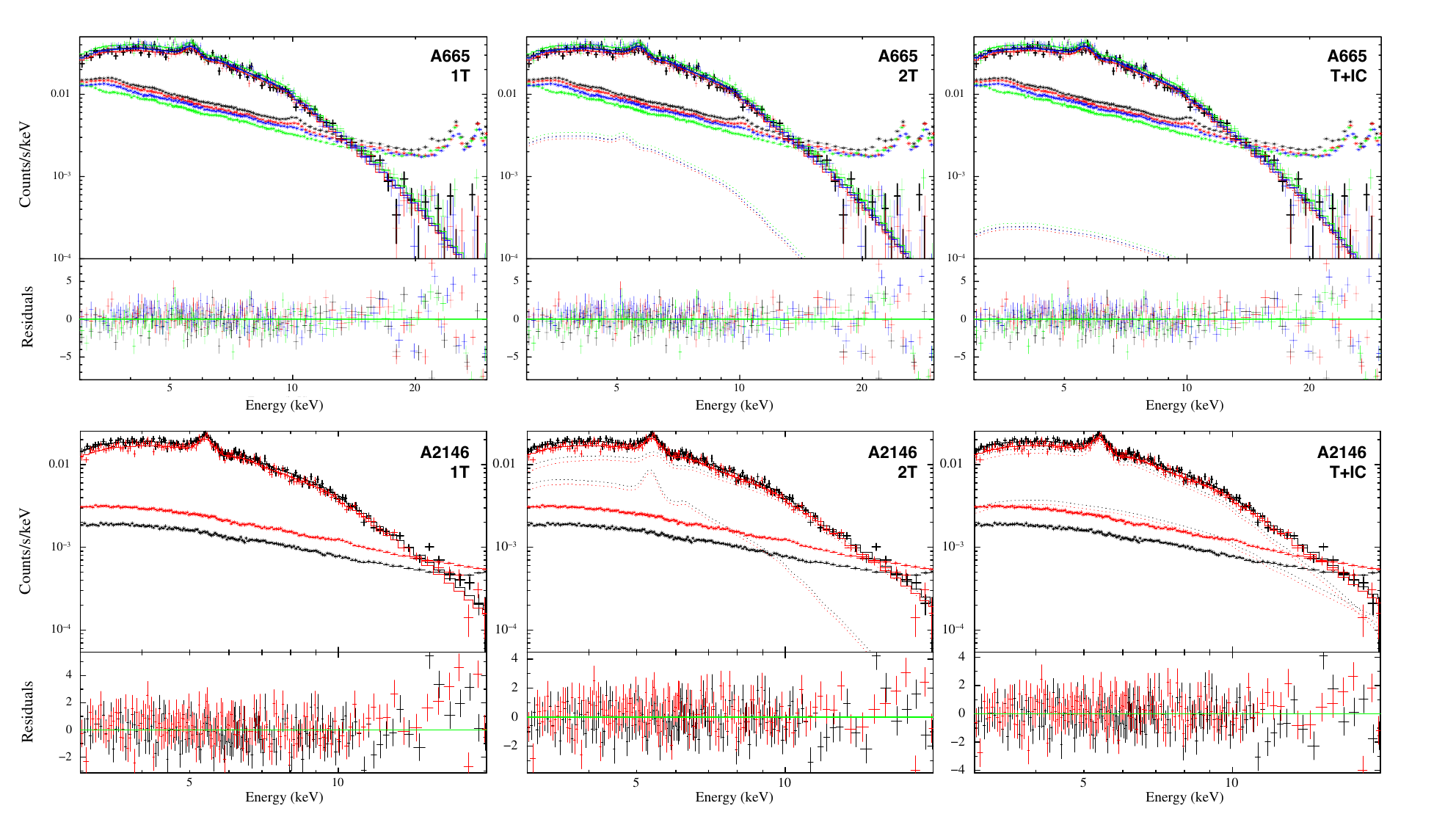}
\caption{Global fits to the A665 (\textit{upper panels}) and A2146 spectra (\textit{lower panels}) with 1T (\textit{left panels}), 2T (\textit{middle panels}), and T+IC (\textit{right panels}) models. 
For the A665 spectra, black indicates FPMA and red indicates FPMB for Obs. ID 0201002002, whereas for Obs. ID 0201002002, FPMA and FPMB are indicated by green and orange, respectively. For the A2146 spectra, black indicates FPMA and red indicates FPMB. For both A665 and A2146, crosses show the data and the background is denoted with asterisks. The dashed curves correspond to the model components to visualize their contributions to the composite model. For plotting purposes, adjacent bins are grouped to ensure a detection significance of at least 10$\sigma$, with maximum 20 bins.}
\label{fig:GlobalSpectra}
\end{figure*}

Spectra were extracted using the regions shown in Figure~\ref{fig:region} with {\tt nuproducts} and fit using {\tt XSPEC}. The fitted spectra for each of the clusters are shown in Figure~\ref{fig:GlobalSpectra}.
In both clusters, the FPMA and FPMB spectra vary negligibly. For A665, we jointly fit both observations, tying together each of the model parameters, except for model normalizations. 
All our spectra are grouped by 30 counts per channel. We use the modified Cash statistic ({\tt statistic cstat} in {\tt XSPEC}) to find best-fit parameters for each model \citep{Cstat}, as this allows us to cut down on the time to fit the data while avoiding loss of information \citep{rojas21}. Additionally, the Cash statistic does not bias our result like $\chi^{2}$ would as we are working with Poissonian data. For A2146, all spectra were fit between 3 and 20 keV instead of 30 keV due to difficulties fitting the data above 20 keV.





\subsection{Models} \label{subsec:models}
    In clusters with radio emission in the form of radio relics and radio halos like A665 and A2146, IC emission must be present, since the two necessary ingredients---relativistic electrons and the CMB---are known to be inhabit the ICM \citep{Liang}. When the IC emission is weak with respect to the thermal emission, as may be the case with these clusters, the model we select to fit the data needs to be able to separate the different types of emission. We used three different models to attempt to characterize the emission of these clusters: a single temperature (1T), a two temperature (2T), and a single temperature plus a power law (T$+$IC) \citep{2014A&A...562A..60O, wikBullet, rojas21}. The thermal emission is represented by the {\tt APEC} model (AtomDB version 3.0.9) within {\tt XSPEC}, which contains as parameters the temperature, abundance, redshift, and normalization.  We allow the metal abundances to be free during the fitting process and we use the abundance table {\tt wilm} \citep{Wilm}. Additionally, as shown in \citet{rojas21}, we can ignore including a foreground absorption model like {\tt phabs} or {\tt tbgas} due to \nustar's effective area not being sensitive to energies below 3~keV. We also allowed the redshift to be free, which changed the redshifts from 0.189 to 0.201 for A665, and from 0.232 to 0.259 for A2146. Details about why leaving the redshift free results in a better fit are discussed in Appendix~\ref{app:GAIN}. The results of our fits are discussed are presented in Table~\ref{table:T1} and discussed in detail in the following sections.

\subsubsection{Single Temperature} \label{subsubsec:T}
    In a merging cluster scenario, one would expect spatial temperature variations across the cluster due to the merger disturbing the gas. While most likely not resulting in the best fitting model to characterize the emission from the cluster, we can still gain some information from fitting a 1T model to the data, namely the average temperature of the galaxy cluster.

    Previous work using \Ginga~data gave an average cluster temperature of $8.26^{+0.95}_{-0.81}$~keV for A665 \citep{Ginga}. 
    Our 1T fit using \nustar\ results in $kT = 9.15\pm 0.1$~keV (statistical uncertainty only), within range of the \Ginga\ result but more precise. The C-stat value for this fit is 2945 with 2681 degrees of freedom (dof). For A2146 a temperature of $7.5\pm0.3$~keV was measured using \chandra\ in \citet{2146Shock}. Our best fit temperature came out to be $kT = 8.29\pm 0.1$~keV, slightly higher than the \chandra\ temperature. The 2.5$\sigma$ disagreement is expected since the merging cluster is far from isothermal \citep{Helen2146} and \nustar\ has a harder response than \chandra. The C-stat value for this fit is 871 with 841 dof.

\subsubsection{Two Temperature} \label{subsubsec:TT}

    Truly determining the temperature structure of merging clusters is difficult. We know that in galaxy cluster mergers the gas is typically distributed non-isothermally, such as in Abell 2163 \citep{bourdin2163} and in A2146 \citep{Helen2146}. Past attempts at measuring the temperature structure in merging galaxy clusters are often biased by which energy bands are favored by a telescope's effective area, calibrations, and projection effects. With \nustar, so long as there are not large amounts of features within the thermal continuum, we can use a 2T model to characterize non-isothermal gas while also taking into account that the telescope is weighted towards hotter temperatures \citep{rojas21}.

    In A665, we find the higher ($T_{\rm H}$) and lower ($T_{\rm L}$) values for the two temperature components to be $T_{\rm H} = 9.34^{+2.9}_{-1.6}$ and $T_{\rm L} = 4.47^{+4.3}_{-2.6}$~keV, with a C-stat value of is 2933 with 2679 dof, suggesting a slightly better fit than the 1T model. In A2146, $T_{\rm H} = 10.9^{+1.7}_{-1.8}$ and $T_{\rm L} = 4.26^{+1.4}_{-1.3}$~keV and the C-stat value is 860 with 839 dof.\\

\subsubsection{T$+$IC} \label{subsubsec:TIC}   
     In the case where a significant portion of the galaxy cluster emission is in the form of non-thermal, IC emission, the 2T model would likely show that with an unphysically hot $T_{\rm H}$. In this scenario, it would be expected that the T+IC model would better fit the spectrum. While containing a single temperature component, the overall shape of the spectrum, especially at the highest energies, would be well fit by {\tt XSPEC}'s power law component that models the non-thermal emission within our bandpass.

    For both clusters, we adopted a photon index value of $\Gamma = 2$, as \citet{2004A&A...423..111F} and \citet{2146Radio} measured a radio spectral index of $\alpha \sim 1$. We do not allow the photon index to be free, as previous similar work with Abell 2163 has shown that photon index will behave like the $T_{\rm L}$ component in the 2T model \citep{rojas21}.
    Our temperature and the power law normalization parameters were left to be free.\\
    \indent In A665 and A2146, the best-fit temperatures obtained from our T+IC model are $kT = 9.12\pm0.2$~keV and $kT = 7.3\pm0.3$~keV respectively, comparable with the single temperature model. The similar temperatures make sense because there are high energy photons that the hard non-thermal component accounts for. To obtain an estimate of the upper limit for the non-thermal flux, we used the upper limit on the confidence range for the power law normalization. From this, we estimated the power law flux from 20--80~keV for the T$+$IC model to be $F_{\rm NT} < 0.595 \times 10^{-12}$~\flux~and $F_{\rm NT} < 0.85 \times 10^{-12}$~\flux~with a $90\%$ confidence level for A665 and A2146 respectively. The C-stat values for these models are 2938 with 2680 dof and 868 with 840 dof, which means that these models don't fit the data as well as the 2T models do. These results imply that the emission from these clusters is purely thermal in nature.

    Theoretically, a 2T+IC model would provide the most stringent constraints on both thermal and non-thermal emission from a merging galaxy cluster.
    In practice, however, the parameters are not well constrained, and either the second temperature component or the IC component becomes suppressed or adjusts to fit a few errant residuals, usually at the lowest or highest energy ends of the spectrum or around the Fe complex.  This behavior results in a model effectively equivalent to either the 2T or T+IC.  Thus, the uncertainty on the power law flux in the 2T+IC case ends up being equivalent to that in the T+IC model. When fitting such a model, the best fit temperature for one of the two thermal components becomes unphysically low and in essence becomes a T+IC model. In this case, the low temperature component is making small corrections at the lowest energies in the spectrum, where the instrument is somewhat less well-calibrated. When forcing both temperature components to be present, the IC emission becomes artificially suppressed based on the way in which the temperature components are constrained. The T+IC model, while more conservative, removes all the assumptions needed to be made to get a lower limit from the 2T+IC modelling. Thus the addition of an extra thermal component provides no advantage over a single temperature and power law model, which can already account for average thermal emission.
        
    In all of the spectra, the residuals at higher energies, where the spectrum is mostly background, have small error bars compared to the larger scatter. This is because the error bars are purely statistical and computed assuming that the background models used are accurate. In the 20 keV regime where the residuals show a large scatter they are underneath the background and not as accurate due to the imperfect modelling of the fluorescent lines here. The systematic error, typically on the order of several percent, is not well characterized and can affect these data points. The scatter of these residuals would be more in line with the size of their error bars if the systematic uncertainty, which hasn't been quantified in detail, were included in quadrature.

\subsubsection{Preferred Model Including Systematic Uncertainties} \label{subsubsec:best}

\begin{figure}
\centering
\includegraphics[scale=0.5]{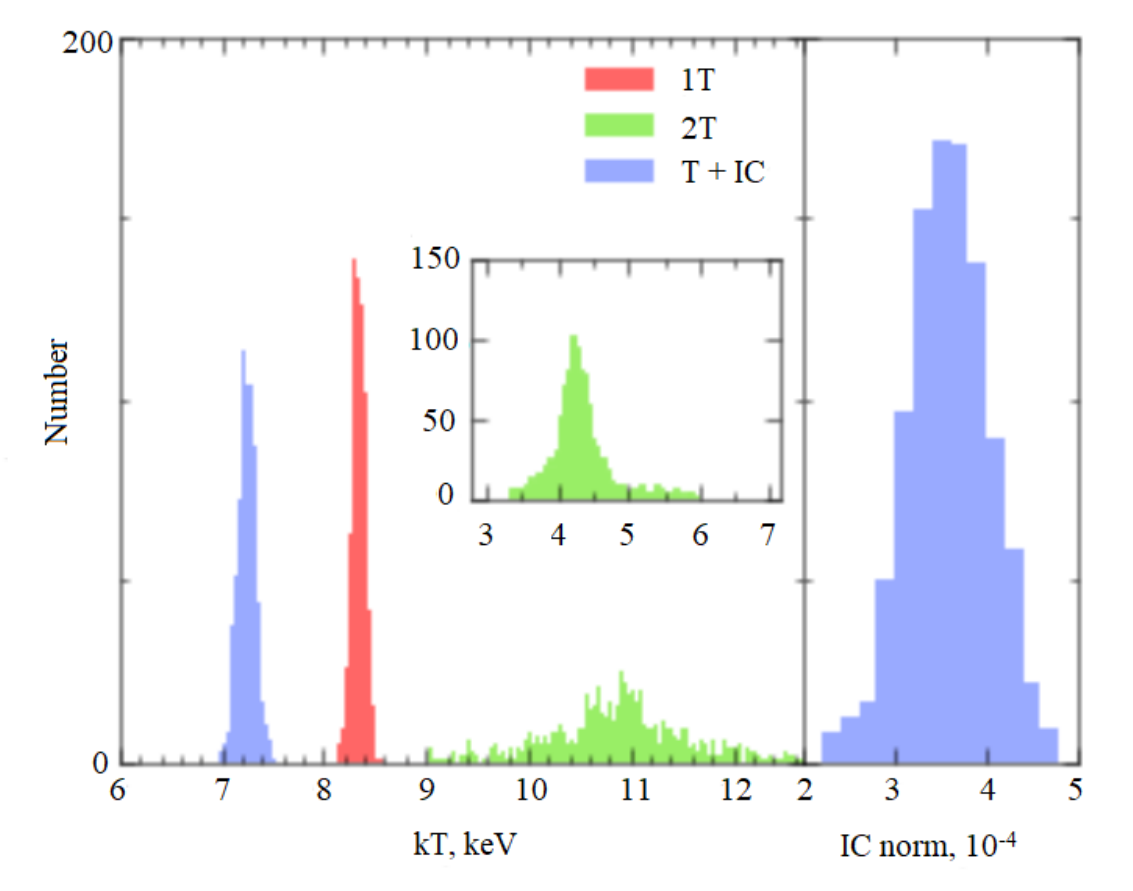}
\includegraphics[scale=0.60]{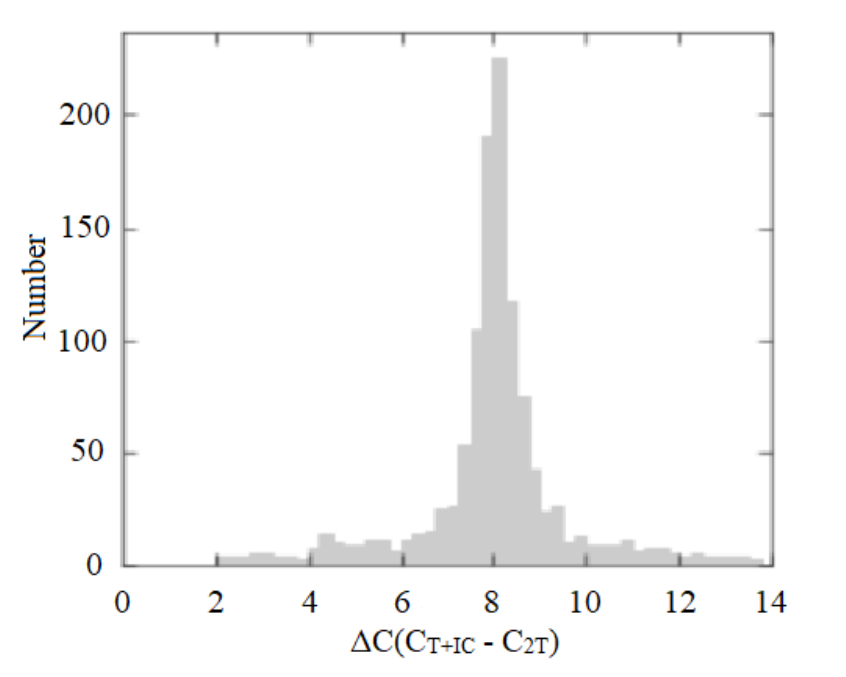}

\caption{Top: The distribution of best-fit parameter values for the 1T (red), 2T (green), and T$+$IC (blue) models using the 1000 realizations of the background (as described in Section~\ref{subsec:uncertainties}) for A2146. Parameters shown are the temperatures of each model and the IC norm. Bottom: The distribution of the difference in C-stat values ($\Delta$$C$) between the T$+$IC and 2T models from fits using the 1000 realizations of the background. The 2T model is preferred in all iterations for describing the \nustar-observed spectra over the T$+$IC model, with the latter not having any realization favoring it over the 2T model. Therefore, we conclude that the data clearly disfavor the addition of a non-thermal component.}
\label{fig:histograms}

\end{figure}

\begin{figure}
\centering
\includegraphics[scale=0.5]{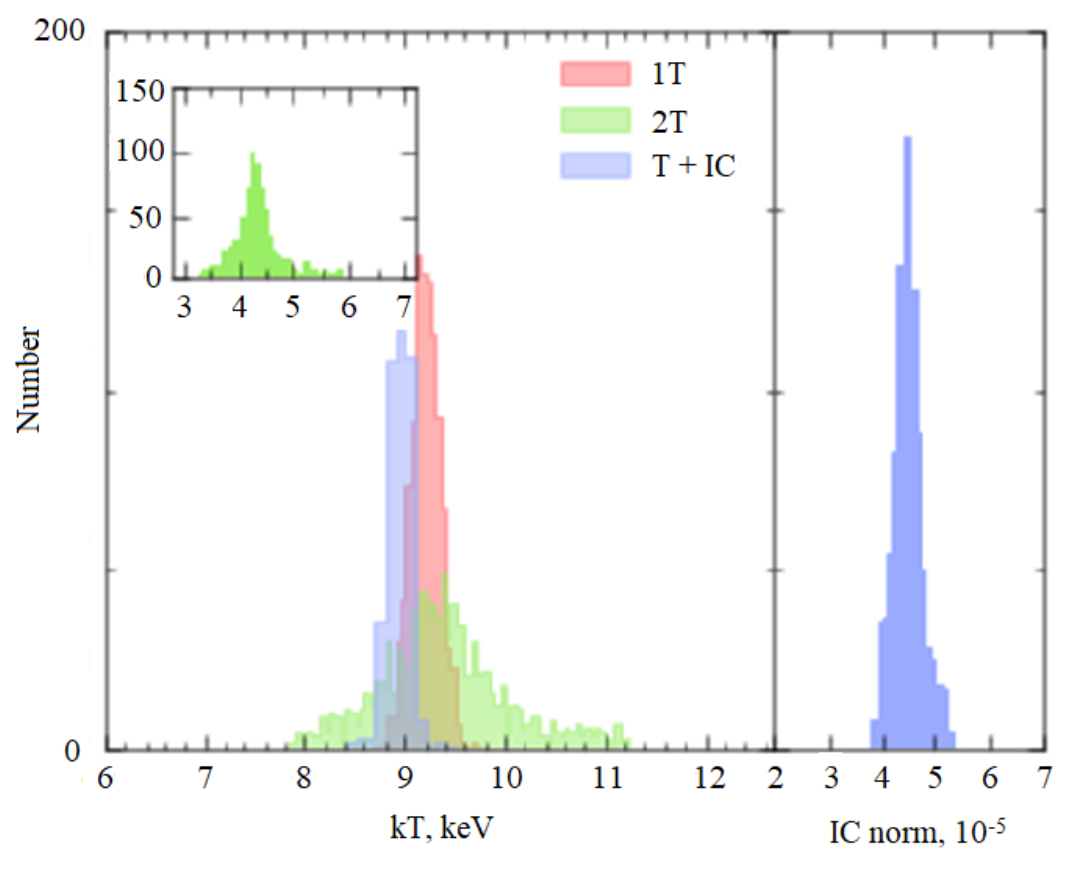}
\includegraphics[scale=0.60]{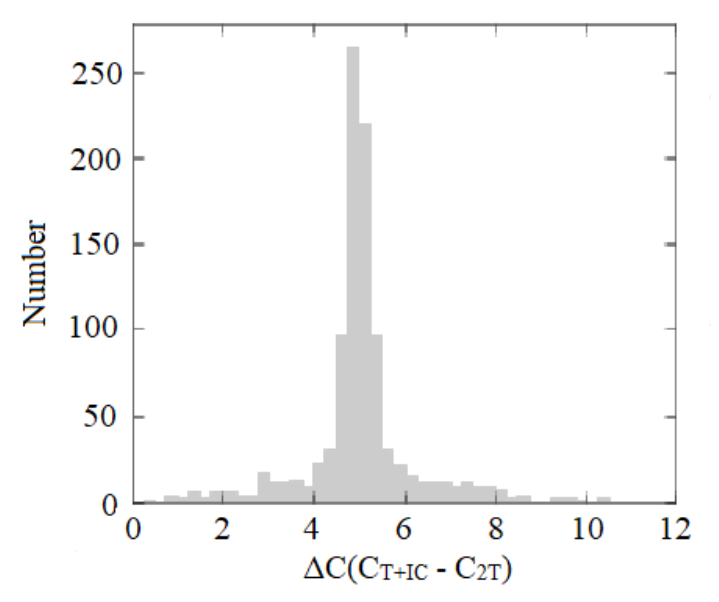}

\caption{Top: The distribution of best-fit parameter values for the 1T (red), 2T (green), and T$+$IC (blue) models using the 1000 realizations of the background (as described in Section~\ref{subsec:uncertainties}) for A665. Parameters shown are the temperatures of each model and the IC norm. Bottom: The distribution of the difference in C-stat values ($\Delta$$C$) between the T$+$IC and 2T models from fits using the 1000 realizations of the background. The 2T model is preferred in all iterations for describing the \nustar-observed spectra over the T$+$IC model, with the latter not having any realization favoring it over the 2T model. Therefore, we conclude that the data clearly disfavor the addition of a non-thermal component.
}

\label{fig:statistics}

\end{figure}

We summarize the results of the previous section in Table~\ref{table:T1}. What we saw from the different model C-stat values was that for both A665 (2933 with 2679 dof) and A2146 (860 with 840 dof), the 2T model best describes the data. This is consistent with previous temperature measurements of the clusters that show non-isothermal gas distributions \citep{665ShockDasadia, Helen2146}. While the nominal C-stat values suggest the 2T model as the best-fitting model, we still have not taken into account background systematics that can greatly affect our models.  \\  
\indent It is well understood that the various background components \nustar\ observes have varying degrees of systematic uncertainties associated with them (see Appendix~\ref{subsec:uncertainties}). To model the systematic errors, we created a distribution of best-fit temperatures for each model by randomly generating 1000 realizations of the background and then fitting the data again. Each new realization of the background had the normalizations of these three components shifted randomly, following a Gaussian distribution within the range of their systematic uncertainty. The distributions of the new best-fit parameters for A665 are shown in Figure~\ref{fig:histograms} and for A2146 in Figure~\ref{fig:statistics}.\\
\indent The 1T model's shape is dependent solely on the temperature, meaning that any shifts in the background will have minimal effects on the temperature. The small change in temperature can be seen by the narrow width of the red histograms in Figures~\ref{fig:histograms} and \ref{fig:statistics}, where the systematic errors end up being comparable to the statistical errors. In the 2T and T+IC model scenarios, the extra parameter introduced will increase the effects of the background systematics within the model. \\
\indent With the inclusion of background uncertainties in the 2T model, we see a much greater effect on the temperature parameters, primarily $T_{\rm H}$. A lower background, for example, when compared to the nominal values results in a higher $T_{\rm H}$ since the spectrum is now turning over at a higher energy and vice versa. The $T_{\rm L}$ temperature then adjusts along with the $T_{\rm H}$ temperature, either increasing or decreasing to correct the low energy portion of the spectrum. The much more evident effects on the background on this model can be seen in the green histograms in Figures~\ref{fig:histograms} and \ref{fig:statistics}. \\
\indent The T+IC model behaves slightly differently than the 2T model when we include background systematics. The temperature parameter of this model is similar to the 1T model in that it is responsible for the shape of the spectrum. In this scenario, it is instead the power law normalizations that vary more with background changes. Any change to the normalization of the power law due to shifts in the background results in a change to the overall IC flux measured, as the IC flux closely resembles the shape of the background. What we observe is that there are only small variations despite sometimes large changes to the background. Were there significant amounts of IC emission in these clusters, we would clearly see it here as the background would have great effects at higher energies. What we are seeing is that the T+IC model actually begins to behave like the $T_{\rm L}$ parameter in the 2T model, where the IC flux behaves like a thermal component that corrects the model at low energies. These results are presented in the blue histograms in Figures~\ref{fig:histograms} and \ref{fig:statistics}. 

With our analysis so far, we have ruled out any detectable presence of IC scattering assuming our nominal background is the true background. In reality, the true background may vary by some amount from the nominal background (see Appendix~\ref{app:BGD} for more details). These variations must be taken into account in order to fully rule out the possibility of IC emission.
To randomly vary the background components, we used the systematic uncertainties described in Appendix~\ref{app:BGD} as the standard deviation and then used the new backgrounds to create best fit models and compare them using their C-stat values. What we find is that even after taking into account systematic uncertainties, the 2T model is always preferred for describing the spectra over the T+IC model. This is reflected in our C-stat distribution histogram in Figures~\ref{fig:histograms} and~\ref{fig:statistics}. While the difference in C-stat values varies somewhat, it is centered around the nominal values of 8 for A2146 and 5 for A665, with no iterations suggesting a T+IC detection.

Out of our three models, we have found that the 2T model best describes the spectra of both A665 and A2146. This does not, however, tell us if the 2T model is the best fitting model for the data. The magnitude of the C-stat solely depends on the number of bins used and the data values, so it does not provide any information about the goodness-of-fit for a model to a set of data \citep{Ftest}. It could still be the case that all of these models are poor fits to the data. To rule this out, we use the {\tt ftest} in {\tt XSPEC} to quantify how reasonable it is to add an extra model component from the T+IC model to the 2T model. We find F-test probabilities of 0.033 and 0.005 for A665 and A2146, meaning that there is a 97.3\% and 99.5\% chance, respectively, that the 2T model is truly the better model to fit the data compared to the T+IC model.  Of course, the ICM of any cluster has a multi-temperature structure, but the spectral resolution and statistical quality of our data allows a 2T model to describe it adequately.

\begin{deluxetable*}{ccccccccc}
\tabletypesize{\scriptsize}
\tablewidth{0pt}
\tablecaption{This table contains the results of our fits using the 1T, 2T, and T+IC models for A665 and A2146. The redshift for all fits was allowed to be free. Nominally, the redshifts for A665 and A2146 are 0.189 and 0.232 respectively. When allowed to change, they became 0.201 and 0.259. See Appendix \ref{app:GAIN}~for more information. Errors are presented as statistical followed by systematic.}
\label{table:T1}

\tablehead{
& & \colhead{Temperature} & \colhead{Abundance} & \colhead{Norm\tablenotemark{a}} & \colhead{$kT$ or $\Gamma$} & \colhead{Norm or IC Flux\tablenotemark{b}} & & \\
\colhead{Galaxy Cluster} & \colhead{Model} & \colhead{(keV)} & \colhead{(Solar)} & \colhead{($10^{-2}$~cm$^{-5}$)} & \colhead{(keV or ...)} & \colhead{($10^{-2}$ or $10^{-12}$~erg~s$^{-1}$~cm$^{-2}$)} & \colhead{C-stat\tablenotemark{c}} &
\colhead{dof}
}
\startdata
Abell 665 & 1T & $9.15 \pm 0.1, 0.3$ & $0.39 \pm 0.03, 0.03$ & $1.6 \pm 0.1, 0.1$ & ... & ... & $2945^{+131}_{-124}$ & 2681 \\
& 2T & $9.34^{+2.9, +1.9}_{-1.6, -2.1}$  & $0.40^{+0.13, +0.12}_{-0.19, -0.08}$ & $1.1^{+0.2, +0.4}_{-0.4, -0.3}$ & $4.47^{+4.3, +1.8}_{-2.6, -1.7}$ & $0.6^{+0.6, +0.3}_{-0.3, -0.2}$ & $2933^{+129}_{-132}$ & 2679 \\
& T+IC & $9.12 \pm 0.2, 0.4$  & $0.40 \pm 0.04, 0.02$ & $1.1 \pm 0.2, 0.1$ & 2 (fixed) & $0.468^{+0.127, +0.171.}_{-0.134, -0.103}$ & $2938^{+121}_{-125}$ & 2680 \\
Abell 2146 & 1T & $8.29 \pm 0.1, 0.1$  & $0.53 \pm 0.06, 0.04$ & $5.6 \pm 0.3, 0.2$ & ... & ... & $871^{+109}_{-114}$ & \phn841 \\
& 2T & $10.9^{+1.7, +2.1}_{-1.8, -1.9}$  & $0.58^{+0.08, +0.14}_{-0.11, -0.07}$ & $3.4^{+0.6, +0.5}_{-0.4, -0.3}$ & $4.26^{+1.4, +1.9}_{-1.3, -1.2}$ & $2.8^{+0.7, +0.5}_{-0.5, -0.6}$ & $860^{+118}_{-122}$ & \phn839 \\
& T+IC & $7.3 \pm 0.3, 0.2$  & $0.63^{+0.04, +0.08}_{-0.8, -0.10}$ & $4.7 \pm 0.2, 0.3$ & 2 (fixed) & $0.600^{+0.251, +0.132}_{-0.212, -0.114}$ & $868^{+115}_{-111}$ & \phn840 \\
\enddata
\tablenotetext{a}{Normalization of the {\tt APEC} model, given by $(10^{-14}/[4 \pi (1+z)^{2} D^{2}_{A}]) \int n_{e} n_{H} dV$ where $z$ is the redshift, $D_{A}$ is the angular diameter distance, $n_{e}$ is the electron density, $n_{H}$ is the ionized hydrogen density, and $V$ is the volume of the cluster.}
\tablenotetext{b}{20--80~keV}
\tablenotetext{c}{Distribution of C-stat values from the 1000 realizations shown.}
\end{deluxetable*}


\section{Summary and Discussion} \label{sec:Summary}

\nustar\ observed A665 twice, once for a time period of 97 ks and the second time for a period of 91 ks, which were then cleaned down to 91 ks and 85 ks after a manual filtering process. The same was done for A2146, which was observed for a period of 285 ks and reduced to 255 ks. After modelling and subtracting the background from our spectra, the data for both clusters was modelled with 1T, 2T, and T+IC models. Allowing for systematic background uncertainties, we showed that the 2T model was the best fitting model and also ruled out the presence of a large IC flux for both clusters. 

\subsection{Non-thermal Emission}

\begin{figure}[h]
\centering

\includegraphics[scale=0.62]{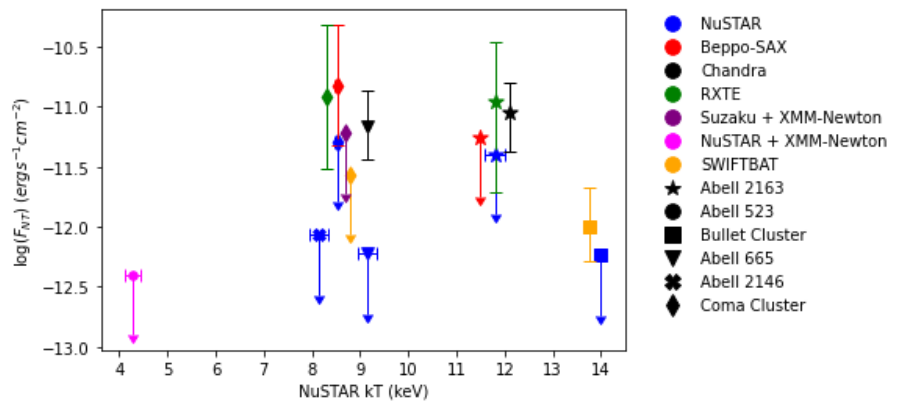}

\caption{This figure shows the 20--80 keV non-thermal flux for 6 clusters as measured by \nustar~(blue), \beppo (red), \chandra\ (black), \RXTE~(green), \Suzaku+\xmm~(purple), \nustar+\xmm~(magenta), and \Swift~(orange). Results for Abell 2163 (star) are taken from \citet{rojas21}, \citet{MillionIC}, \citet{2001A&A...373..106F}, and \citet{2163RXTE}. The Abell 665(downward triangle) flux is also from \citet{MillionIC}. The Bullet Cluster (square) measurements are from \citet{wikBullet} and \citet{2010ApJ...725.1688A}. Coma Cluster (diamond) results are taken from \citet{FabioComaNuStar}, \citet{FF665Beppo}, \citet{Rephaeli665RXTE}, \citet{wikSuzaku}, and \citet{wikSWIFT}. Abell 523 (circle) results are from \citet{Abell523}. All data points are plotted with the \nustar~measured temperature, although some have been slightly shifted for clarity.}

\label{fig:history}

\end{figure}

The 90\% upper limits 20--80~keV flux of non-thermal emission coming from A665 and A2146 is $F_{\rm NT} < 0.595 \times 10^{-12}$~\flux\ and $F_{\rm NT} < 0.85 \times 10^{-12}$~\flux\ respectively.

From our T$+$IC model we can set 90\% upper limits on the 20--80~keV flux of non-thermal emission coming from A665 and A2146 of $F_{\rm NT} < 0.595 \times 10^{-12}$~\flux\ and $F_{\rm NT} < 0.85 \times 10^{-12}$~\flux\ respectively. Based on the statistical tests and comparisons described in Section~\ref{subsubsec:best}, we can safely rule out the presence of large amounts of IC scattering when comparing the 2T and T+IC models. The data is best fit by a two temperature component, purely thermal model. The variation in power law normalizations shown in the right panel of Figures~\ref{fig:histograms} and \ref{fig:statistics}, which translate to variations in IC fluxes, is due to the IC flux acting as an extra, lower temperature component in the T+IC model. The addition of this power law component does not fit the data as well as the 2T model as shown in C-stat comparisons the bottom panels of the previously mentioned figures. It should be noted that the C-stat difference is lower for A665 most likely due to the faintness of the cluster.\\
\indent For A665, our IC flux upper limit is an order of magnitude lower than the one found by \citet{MillionIC} (black triangle in Figure~\ref{fig:history}). Although they agree that a single temperature model does not entirely describe the cluster emission, they conclude that the temperature and metallicity variations across the cluster play an important role in the detection of non-thermal-like components in spectral fits. \citet{SandersPerseus} used the same calibration files for Chandra as used by Million in the aforementioned paper, where they also found evidence for non-thermal like emission in Perseus. It was later pointed out by \citet{MolendiPerseus} that errors in the effective areas used in those calibration files inflated the apparent significance of the non-thermal component in their models.  

In addition, due to the susceptibility of \chandra\ to galactic foreground, variations in the foreground column density of atomic gas, and an additional power law cannot be easily disentangled (see \citet{rojas21} for more details). 

For A2146, there are no previous IC flux measurements that have been published.

Including A665 and A2146, there are now a total of six published IC flux upper limits using \nustar~as of this work. Each of these six clusters that have been studied have various other upper limits or detections done with several other observatories, such as \chandra, \Suzaku, \RXTE, \beppo, \Swift, and \xmm. While no one has compiled IC flux measurements for \nustar~in the same way that \citet{OtaReview} did for \Suzaku, we wanted to design a plot similar to \citet{2014A&A...562A..60O} for \nustar~measurements compared to the other observatories. Figure~\ref{fig:history} shows a plot of the non-thermal fluxes for the six clusters as a function of the \nustar~gas temperature. The overall trend that can be seen in the figure is that \nustar~has consistently provided the most stringent constraints on the IC flux as well as only providing upper limits and no detections. It is important to note that the fluxes reported using \chandra~were all provided in the 0.6--7 keV range and the \nustar~and \Swift~Bullet Cluster fluxs from 50--100 keV and 20--100 keV respectively. These have all been converted to 20--80 keV in the plot for the sake of comparison between the rest of the measurements. It is also important to note that estimating the flux depends on the various assumptions made by each author, such as how the thermal component was modelled, what power law index was chosen for the non-thermal component, different apertures, and different extraction regions for spectra, among other things.

\subsection{Cluster Magnetic Field} \label{subsec:magnets}
    As descibed in Section~\ref{sec:intro}, with an upper limit on the IC flux, we can set a lower limit on the average magnetic field strength $B$ using Equation~\ref{eq:2}. A total diffuse radio flux of 43.1 mJy inside our global extraction region was determined from VLA observations at 1.4~GHz for A665 \citep{Vacca665} and 1.5 mJy for A2146 \citep{2146Radio}.

With the T$+$IC model we obtain lower limits of 0.14~$\mu$G and 0.011~$\mu$G for A665 and A2146, respectively. \citet{Vacca665} obtained an equipartition estimate for the magnetic field in A665 of $B = 1.3~\mu$G. In their work, they calculated the magnetic field strength assuming local equipartition of energy density between relativistic particles and the intracluster magnetic field. To set this condition, the magnetic field energy contributions should equal the relativistic particle contributions. There is another estimate of the magnetic field strength in A665 done by \citet{2004A&A...423..111F}. In their estimate they also apply equipartition and calculate the magnetic field strength to be $B = 0.55~\mu$G. For A2146, there are no available magnetic field strength estimates, likely due to the fact that until recently \citep{2146Radio}, the extended radio emission was difficult to measure \citep{RussellRadio}. 

The magnetic field limit for A2146 is an order of magnitude lower than that of A665 and other clusters with IC upper limits, such as those presented in Figure~\ref{fig:history}, which is simply a result of its weak diffuse radio emission.
As such, it is not an ideal target for an IC search, but since the radio halo was only recently discovered, this work provides the first IC-based limit on the magnetic field in A2146.
The limit in A665 is more typical for IC searches although it still falls below estimates using equipartition arguments.

The importance of \nustar's ability to provide these magnetic field strength limits is that they run counter or rule out past detections of IC emission, which implied low magnetic field strengths. Such lower limits still allow for the possibility of high magnetic field strengths, which could be dynamically important, especially in cluster outskirts. Studies done on Abell 3667 suggest that that strong magnetic field strengths in the cluster outskirts could create a 20--30\% pressure contribution to hydrostatic equilibrium \citep{A3667, A3667NW}.

This extra pressure contribution is currently omitted in many HSE models and may possibly resolve the $\sim$20\% discrepancy between mass measurements done using weak lensing analysis and X-ray measurements \citep{Biffi}.

Rotation measure (RM) synthesis estimates provide another method for measuring the magnetic field in clusters, giving volume-average strengths a few times higher \citep[e.g.,][]{2010A&A...513A..30B}.
However, RM-estimated magnetic fields are weighted by electron density, so correlations between density and field strength could bias the value of $B$.
Lower limits from IC searches thus provide an important constraint on magnetic fields in clusters.

\subsection{Future Work}

In future work, we would like to limit our non-thermal emission search to local regions within the clusters, using \xmm\ data to include soft X-rays and joint-fit them to the \nustar\ data. If the diffuse IC emission is more localized within the ICM, this would provide enough sensitivity to detect it. Additionally, the inclusion of the soft X-ray data may tighten constraints on the measured parameters within our three models presented earlier. Another topic for exploration with these clusters is the detection of a possible non-thermal bremmstrahlung, corresponding with a population of suprathermal electrons \citep{super}. Combining the \nustar\ data with \LOFAR\ radio data could bridge the gap between the thermal and non-thermal electron distributions as well as provide insight into the non-thermal Sunyaev-Zeldovich effect \citep{ntbrem, ntradiation}.

\acknowledgments
This work made use of data from the \nustar\ mission, a project led by the California Institute of Technology, managed by the Jet Propulsion Laboratory, and funded by NASA.
RARB and DRW gratefully acknowledge support from NASA grants NNX17AH31G and 80NSSC18K1638. Basic research in radio astronomy at the Naval Research Laboratory is supported by 6.1 Base funding. 
This research has made use of the \nustar\ Data Analysis Software ({\tt NuSTARDAS}) jointly developed by the ASI Science Data Center (ASDC, Italy) and the California Institute of Technology (USA). VV acknowledges support from INAF mainstream project "Galaxy Clusters Science with LOFAR 1.05.01.86.05".
We thank the referee for useful comments.

\typeout{}
\bibliography{mybibliography}

\appendix 

\section{Background Characterization and Modelling} \label{app:BGD}


\nustar~is unique from the other major X-ray observatories in that it has a large open mast that separates the focal plane from the optics modules. Due to this feature, stray light produces a gradient across the FOV of the observatory. The background caused from this stray light gradient must be modelled and separated from that of the instrumental background. The instrumental background is usually spatially uniform due to it being produced by unknown cosmic background sources with varying intensity. This section will elaborate briefly on how we applied the methods used in \citet{wikBullet} and \citet{rojas21} to characterize the background with {\tt nuskybgd} as well as a description of the individual components of our background models. \\

\begin{figure}
    \centering
    \includegraphics[scale=0.40]{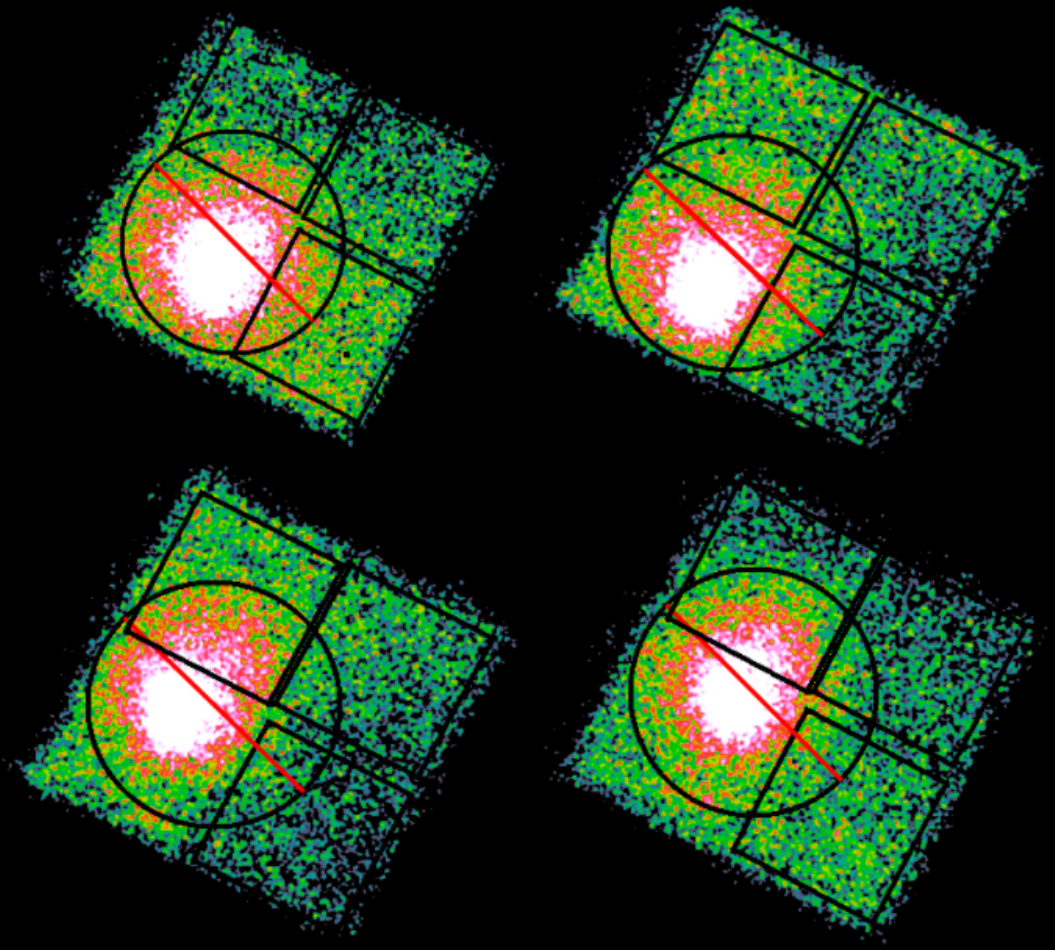}
    \caption{False color images of FPMA (left panels) and FPMB (right panels) for both observations of A665 (OBSID 70201002002 on top, OBSID 70201003002 on bottom). Overlaid are the {\tt nuskybgd} regions used to extract background spectra from each of the four detectors. Note that we did not use a region in DET2 as the cluster's emission is largely present there without much room to make a good enough region to capture background emission. We also used an exclusion region around the cluster to remove emission leaking into our regions. Background fits can be seen in Appendix~\ref{app:BGD}.}
    \label{fig:A665bgd}
\end{figure}

\begin{figure}
    \centering
    \includegraphics[scale=0.35]{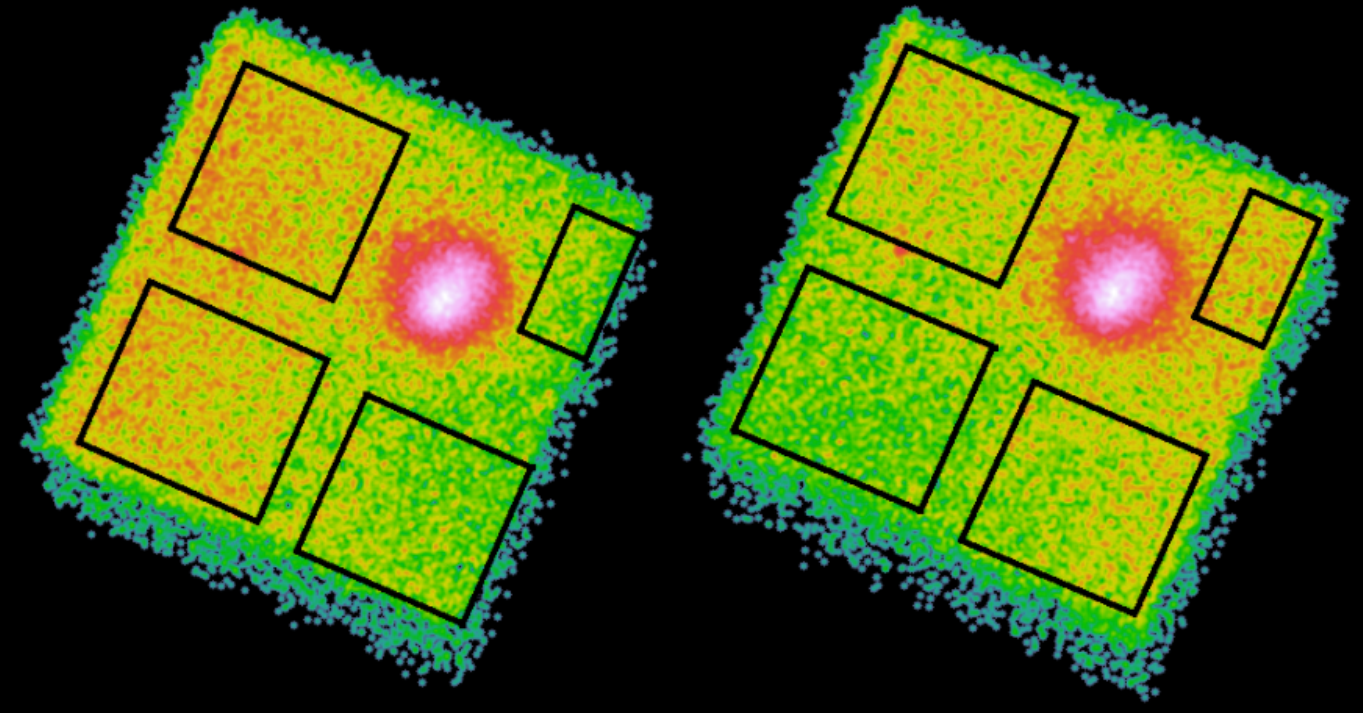}
    \caption{False color images of FPMA (left panel) and FPMB (right panel) for A2146. Overlaid are the {\tt nuskybgd} regions used to extract background spectra from each of the four detectors. Background fits can be seen in Appendix~\ref{app:BGD}.}
    \label{fig:A2146bgd}
\end{figure}

\begin{figure*}
\includegraphics[width=180mm]{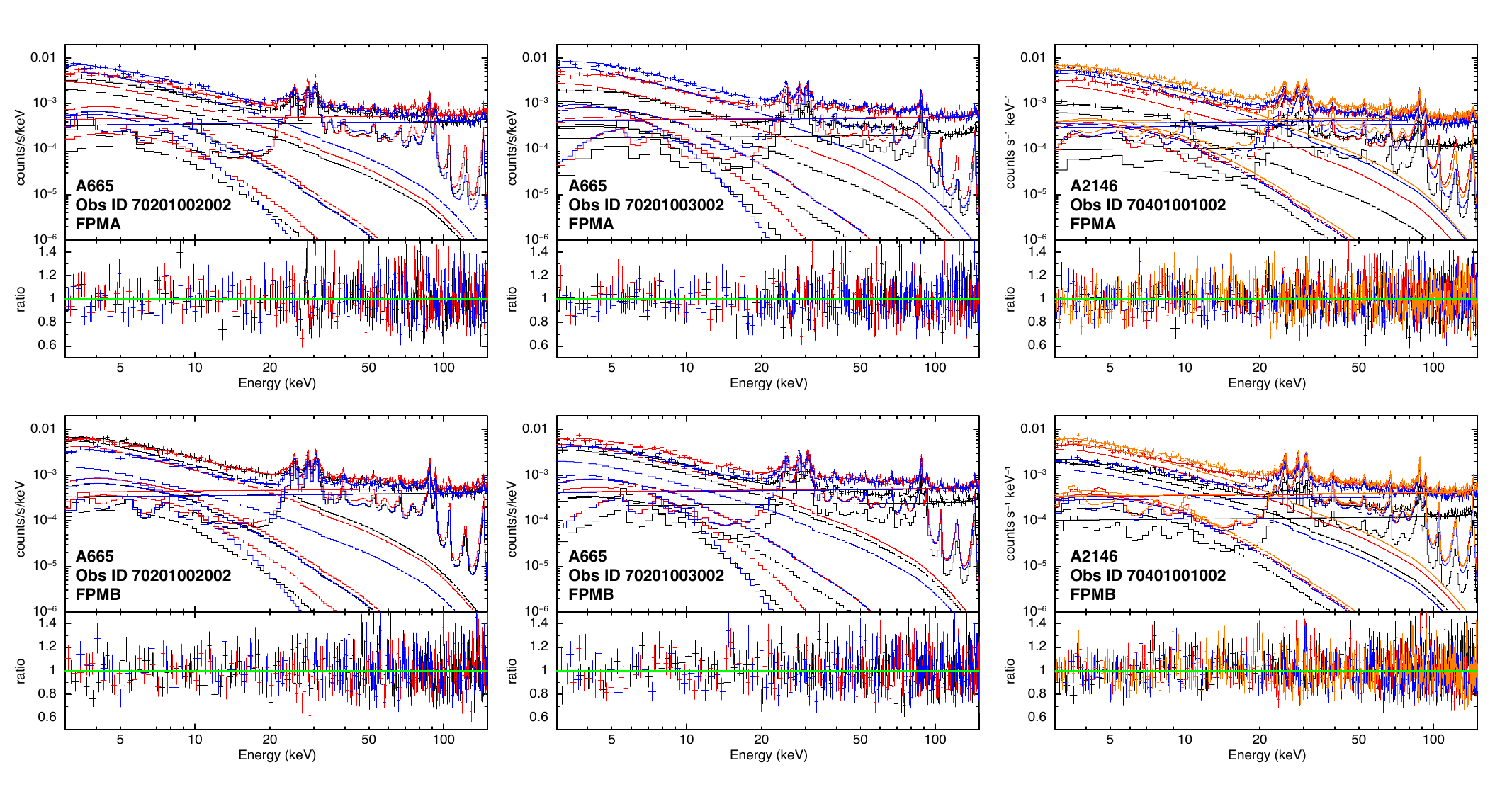}
\caption{
Fits to the background spectra extracted from the regions shown in Figure~\ref{fig:region}. The top and bottom panels show the spectra from FPMA and FPMB, respectively. The order from left to right is A665 (first observation), A665 (second observation), and A2146. Each region is represented by a different color.
Solar contributions produce the background at softer energies ($<$5~keV) which is combined with the model that includes instrumental background lines present at all energies. The undeflected CXB that passes through the aperture stops dominates between 10--20~keV, while the CXB that is focused through the optics (fCXB) is $\sim$10$\times$ fainter. ICM emission from the cluster produces the steeper spectral models in this energy range. At higher energies, cosmic ray-induced activation and fluorescence lines dominate, along with a flat continuum component representing the overall instrumental background level. For plotting purposes, adjacent bins are grouped until they have a significant detection at least as large as 10$\sigma$, with maximum 15 bins.}

\label{fig:bgdspec}

\end{figure*}

\indent There are four components that compose the background: internal, aperture stray light, reflected and scattered stray light, and focused cosmic background \citep{wikBullet}. Since these components are fairly well understood, we can model the composition of them from source-free regions, shown in Figures~\ref{fig:A665bgd} and~\ref{fig:A2146bgd}. We can then take what we know about how the components vary spatially and apply the models to source regions. These regions are never completely source free due to the diffuse nature of the emission from galaxy clusters. To account for this emission, we create a contamination file that contains APEC models with a free temperature parameter used to characterize thermal emission from the clusters. The results for our background fits for the clusters can be seen in Figure~\ref{fig:bgdspec}. We found no notable differences observed in the backgrounds between FPMA and FPMB. In our A2146 results, there is a notable bump around the 6 keV iron line, reflecting a gain issue that we discuss in Appendix~\ref{app:GAIN}. 

\subsection{Background Characterization} \label{subsec:components}


    There are a few different sources that contribute towards the internal, instrumental background of \nustar. The first is due to the radiation environment in which the observatory finds itself in its low Earth orbit. Additionally, activation and fluorescence lines, which dominate the background from 22-32 keV, are present. Finally, the internal background is affected by the Compton scattering of gamma rays that pass through the anti-coincidence shield and strike the detector.

    Previously, we mentioned the open design of \nustar~and how that allows stray light to strike the telescope. Ideally, the aperture stops that are on the focal plane bench would block these unfocused X-rays from coming in. Unfortunately, due to technical limitations, there is a small window of about 3 degrees for these photons to pass through. A smooth gradient is produced from the roughly uniform cosmic X-ray background (CXB) stray light thanks to the optics bench partially blocking the FOV of the detector. This gradient depends on the position of the optics module and the orientation of the detectors. We use the CXB spectral model used by \citet{1987NASCP2464..339B}, the {\tt HEAO1 A2} spectral model, valid from 3--60 keV. 
   
    Another pitfall of the open design of \nustar~is that reflected and scattered X-rays from the entire sky can also strike the detectors. While many parts of the telescope can serve as a reflecting surface, the prime culprit is the backside of the aperture stops. Roughly 10-20$\%$ of the unfocused light that gets through is reflected \citep{wikBullet}. There are three main sources of reflected light: the Sun, the Earth, and the CXB. Terrestrial and solar light are primarily found around 1 keV during high periods of solar activity such as solar flares. 

    The focused cosmic background (fCXB) is the last component that we have to model. This is the contribution to the background from foreground and background sources that are unresolved within our field of view. These are mainly observed below 15 keV and correspond to roughly 10$\%$ of the photons from the CXB, with most of them entering unreflected through the aperture stops \citep{wikBullet}.

\subsection{Systematic Uncertainties} \label{subsec:uncertainties}
    To calculate systematic uncertainties, we follow the analysis detailed in \citet{wikBullet} and \citet{rojas21} for the Bullet Cluster and Abell 2163 respectively. Since a potential IC emission is expected to be faint, it is important to consider possible systematic uncertainties in the background to prevent a false detection and derive accurate upper limits. For the instrumental background, we adopt a 90\% uncertainty of roughly 3$\%$ to account for systematic variations.
        For the aperture CXB, we adopt a systematic uncertainty of 8$\%$, the origin of which is from cosmic variance due to variability in the flux of large-scale structure depending on the solid angle taken. The uncertainty in the fCXB is calculated in similar fashion and results in potential variations of up to 42\%. 
    




\section{Gain Issues with \nustar} \label{app:GAIN}

\begin{figure}[h]
\centering
\includegraphics[scale=0.75]{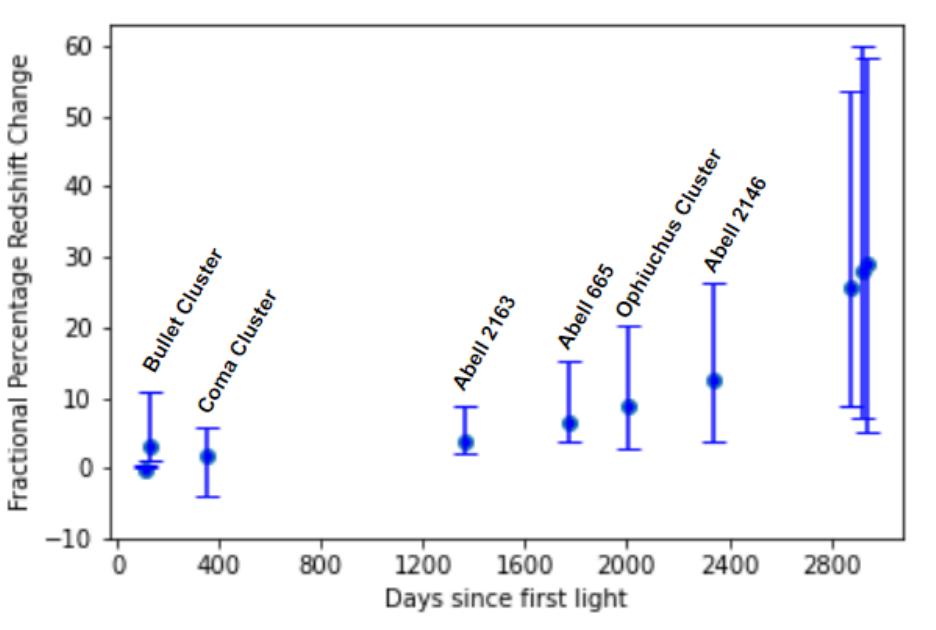}

\caption{This figure shows observations conducted in every \nustar~observation cycle since first light on June 28th, 2012. The first two points are observations of the Bullet Cluster taken pre-cycle 1. The final three points are MACS J0717.5+3745, Abell 2319, and Abell 1795 in that order.}

\label{fig:redshift}

\end{figure}

\indent The \nustar~focal planes each contain a calibration source made of $^{155}$Eu with prominent emission lines at 6.06, 6.71, 86.54, and 105.4 keV. The calibration source is on a movable arm that can move the source into the FOV and can be used to confirm the gain of \nustar~during an observation. The CALDB files contain a FPM gain file that applies a 0.2\% gain shift yearly. While fitting the data, particularly A2146 (a cycle 4 \nustar~target), we found that fixing the redshift to its nominal value of $z \sim0.2323$ did not result in the best fit. Notably, we observed worse residuals at low energies around the 6 keV lines mentioned previously. When allowing the redshift to be a free parameter, these residuals were improved. We believe that the gain shift correction is not being properly applied to the low energy 6 keV lines. We have included a plot showing how the redshift has changed as a function of years \nustar~has been observing, shown in Figure~\ref{fig:redshift}. We want to investigate if this was just a peculiar instance or if it is a problem that is consistent or deteriorating across \nustar\ observations. It is not a major issue, as simply allowing the redshift to be free fixes it. However we wanted to try to track down what the problem is or if we could find a more concrete solution. 

\begin{figure}[h]
\centering
\includegraphics[scale=0.75]{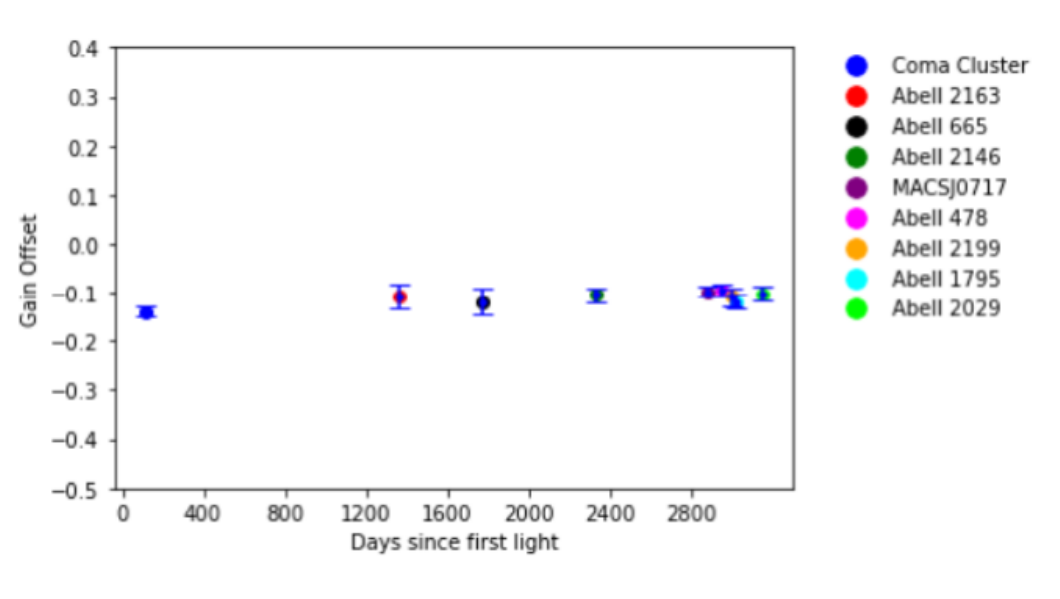}

\caption{This figure shows how the gain offset, as obtained by the methods described in Appendix~\ref{app:GAIN}, varies over the life time of \nustar. The offset remains consistently around a value of 0.11 keV, suggesting that the issue has continuously existed across all observation cycles.}

\label{fig:offset}

\end{figure}

\begin{figure}[h]
\centering
\includegraphics[scale=0.75]{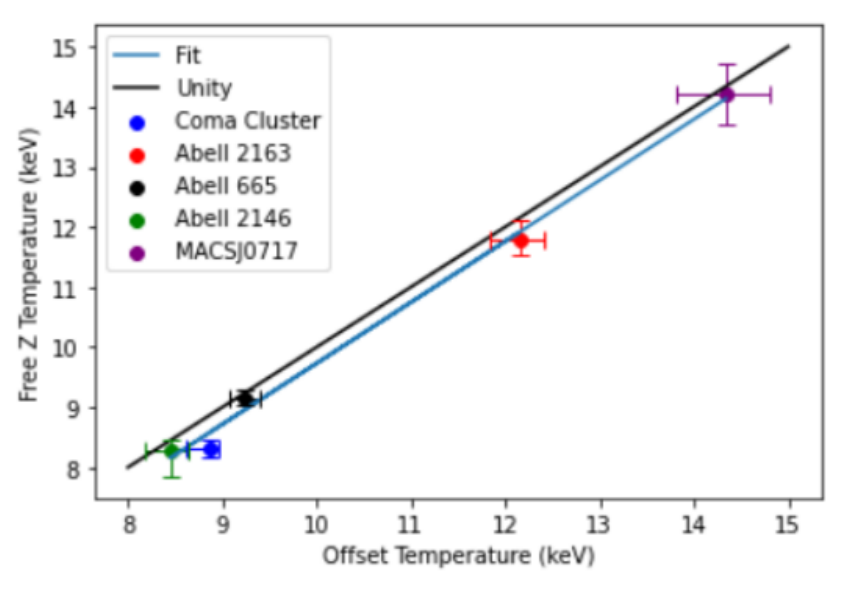}

\caption{This figure shows the free redshift versus offset temperature relation between five clusters. The green line is our best fit line to the data, while the black line is the x=y line showing a 1:1 relation for reference and comparison to Figure~\ref{fig:fixedoffset}. The free redshift temperatures are closer to the offset values than in the other case}

\label{fig:freeoffset}

\end{figure}

\begin{figure}[h]
\centering
\includegraphics[scale=0.75]{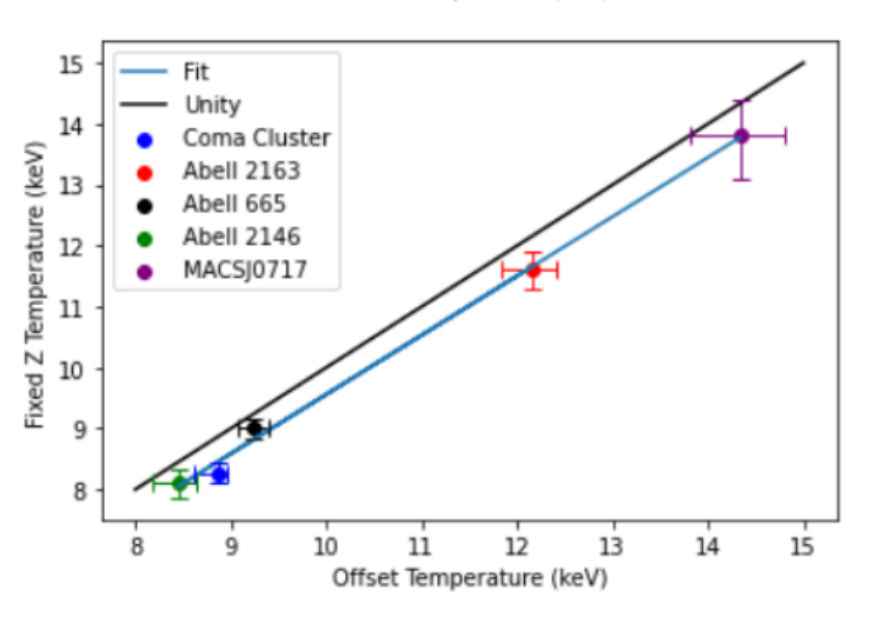}

\caption{This figure shows the fixed redshift versus offset temperature relation between five clusters. The green line is our best fit line to the data, while the black line is the x=y line showing a 1:1 relation for reference and comparison to Figure~\ref{fig:freeoffset}. The fixed redshift temperatures are further from the offset values than in the other case.}

\label{fig:fixedoffset}

\end{figure}

\begin{figure}[h]
\centering
\includegraphics[scale=0.75]{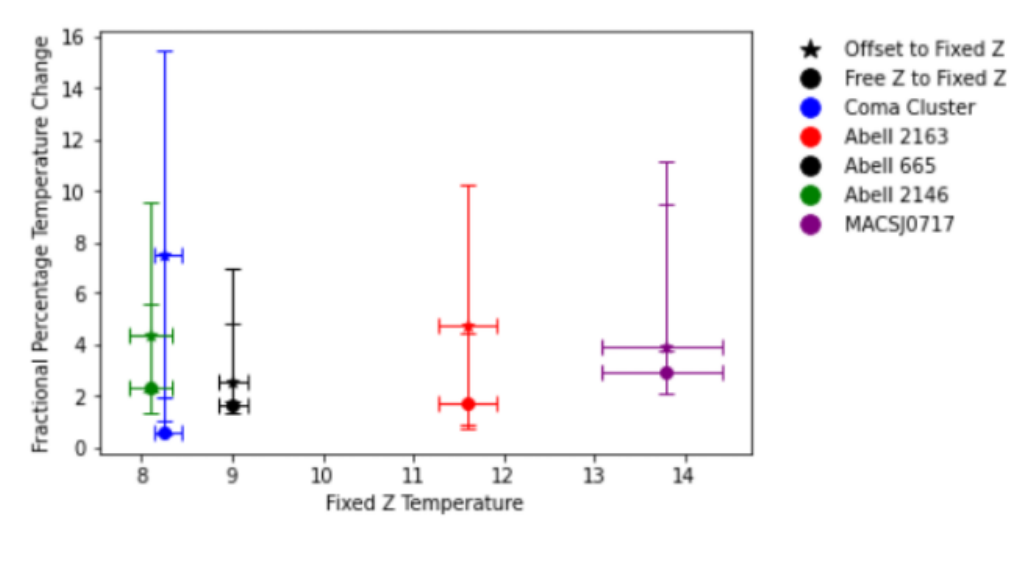}

\caption{This figure shows the fractional percentage temperature change of the same five clusters from Figures~\ref{fig:freeoffset} and \ref{fig:fixedoffset}. There is overlap between the free redshift and offset temperatures, suggesting that either method of correction works.}

\label{fig:percentage}

\end{figure}

    We used the {\tt gain} command in {\tt XSPEC} which modifies the response file gain by shifting the energies that define the response matrix and then adjusting the effective area curve within the loaded ARF to match. The new energy is the original energy divided by the difference between the slope and intercept (E' = E/(slope - intercept)). We fed our model (which was loaded after the initial {\tt APEC} model) an initial gain offset of -0.1 keV while allowing the redshift to be fixed to its original SIMBAD value for each cluster. This was done for the following clusters: Coma, Abell 2163, Abell 665, Abell 2146, MACSJ0717, Abell 478, Abell 2199, Abell 1795, and Abell 2029. The resulting offset values after fitting can be seen in Figure~\ref{fig:offset}. The average offset ended up being $\sim0.11$ keV and there isn't much deviation, suggesting that whatever the issue is it is consistent over time. One thing of note is that across the board, the resulting temperatures including the gain corrected model always gave a slightly higher temperature than the free redshift temperatures. The free redshift temperatures are also always slightly higher than the fixed redshift temperatures. To visualize the difference between the three options, we plotted the offset versus free redshift and offset versus fixed redshift temperatures in Figures~\ref{fig:freeoffset} and~\ref{fig:fixedoffset} respectively. To quantify the difference between the options, we calculated the fractional percentage temperature change which is shown in Figure~\ref{fig:percentage}. The result is that there is a lot of overlap between the free redshift temperature and offset temperature meaning that either method works as a suitable solution. 
    
    Some plausible explanation for these discrepancies have been ruled out already, such as the known issues with DET2 on FPMA\footnote{https://heasarc.gsfc.nasa.gov/FTP/caldb/docs/nustar/cal\_nustar\_20211020/NuSTAR\_Gain\_20211025.pdf}. 
    Not all of the clusters selected here are centered on DET2. Another possible explanation is that the {\tt APEC} model is not properly calibrated. The model was recently recalibrated to fix minor issues after a \Hitomi~observation of the Perseus Cluster and any major issues would have shown up there \citep{Hitomi}. For some clusters like Abell 2146, a fix was to update the CALDB to any post-2019 version. It is likely that a small time-dependent gain shift was fixed in newer CALDB updates. This does not fix the issue in all clusters, however.
    Further investigations on the root cause of this issue are required, since for calibration cases this may be of importance while exploring systematic errors. For everyday science, however, it should not normally matter which of these two methods you select to do. 


\end{document}